\documentclass[nofootinbib,aps,11pt]{revtex4-1}
\usepackage{graphicx}
\usepackage{hyperref}
\usepackage{cancel}
\usepackage{amssymb}
\usepackage{textcomp}
\usepackage{amsmath}
\usepackage{bm}
\usepackage{times}
\usepackage{epsfig}
\usepackage{color}

\begin{document}
\title{\Large Displaced Heavy Neutral Lepton from New Higgs Doublet}
\bigskip
\author{Fa-Xin Yang$^1$}
\author{Feng-Lan Shao$^1$}
\email{shaofl@mail.sdu.edu.cn}
\author{Zhi-Long Han$^2$}
\email{sps\_hanzl@ujn.edu.cn}
\author{Yi Jin$^{2,3}$}
\author{Honglei Li$^2$}
\affiliation{
	$^1$School of Physics and Physical Engineering, Qufu Normal University, Qufu, Shandong 273165, China\\
	$^2$School of Physics and Technology, University of Jinan, Jinan, Shandong 250022, China
	\\
	$^3$Guangxi Key Laboratory of Nuclear Physics and Nuclear Technology, Guangxi Normal University, Guilin, Guangxi 541004, China
}
\date{\today}

\begin{abstract}
Heavy neutral leptons $N$ are introduced to explain the tiny neutrino masses via the seesaw mechanism. For proper small mixing parameter $V_{\ell N}$, the heavy neutral leptons $N$ become long-lived, which leads to the displaced vertex signature at colliders. In this paper, we consider the displaced heavy neutral lepton from the neutrinophilic Higgs doublet $\Phi_\nu$ decay. The new Higgs doublet with MeV scale VEV can naturally explain the tiny neutrino masses with TeV scale $N$. Different from current experimental searches via the $W^\pm\to \ell^\pm N$ decay, the new decays as $H^\pm\to \ell^\pm N$ are not suppressed by the small mixing parameter $V_{\ell N}$. Therefore, a larger parameter space is expected to be detected at colliders. We then investigate the promising region at the 14 TeV HL-LHC and the 3 TeV CLIC. According to our simulation, the DV signature could probe $|V_{\ell N}|^2\gtrsim10^{-19}$ with $m_N<m_{H^+}$, which covers the seesaw predicted value $|V_{\ell N}|^2\sim m_\nu/m_N$.  We could probe $m_{H^+}\lesssim1200$ GeV at the 14 TeV HL-LHC  and $m_{H^+}\lesssim1490$ GeV at the 3 TeV CLIC.

\end{abstract}

\maketitle

\section{Introduction}

Observations of neutrino oscillation mean that the lepton sector of the standard model (SM) should be extended with non-zero neutrino masses \cite{Super-Kamiokande:1998kpq,SNO:2002tuh}. Meanwhile, the cosmological observations require that the sum of neutrino masses should be less than 0.12 eV \cite{Planck:2018vyg}, which indicates that the neutrino masses are below the eV scale. Such tiny neutrino masses can be generated by the Yukawa interaction with SM Higgs doublet as $y \bar{L}\tilde{\Phi}\nu_R$, but the coupling $y\lesssim10^{-12}$ is unnatural small compared with other SM couplings. 

One appealing pathway to explain the origin of tiny neutrino masses is introducing the heavy neutral leptons (HNL) $N$. Through the canonical type I seesaw mechanism ~\cite{Minkowski:1977sc,Mohapatra:1979ia}, the predicted neutrino mass is $m_\nu\sim (y v)^2/m_N$, where $v=246$ GeV is the SM Higgs vacuum expectation value (VEV). For $y\sim\mathcal{O}(1)$, eV scale neutrino mass can be obtained with $m_N\sim\mathcal{O}(10^{14})$ GeV. The success of thermal leptogenesis also requires $m_N\gtrsim10^{9}$ GeV \cite{Davidson:2002qv}. But the supper heavy neutral leptons $N$ are far beyond the reach of current colliders.

If we do not insist on large Yukawa coupling $y$, electroweak scale heavy neutral leptons are allowed by current experimental limits \cite{Abdullahi:2022jlv}. Low-scale leptogenesis is also possible with nearly degenerate heavy neutrino leptons \cite{Akhmedov:1998qx,Pilaftsis:2003gt}. However, the seesaw predicted mixing parameter $V_{\ell N}\sim \sqrt{m_\nu/m_N}\lesssim10^{-6}$ is too small to be tested even in future planned experiments. To study the signature of heavy neutral leptons at  colliders, an electroweak scale $m_N$ with free mixing parameter $V_{\ell N}$ is usually assumed \cite{Han:2006ip}, because testable large $V_{\ell N}$ is theoretically possible in inverse seesaw mechanism \cite{Mohapatra:1986bd,Mohapatra:1986aw}. From the well-known seesaw formula $m_\nu\sim (yv)^2/m_N$, Ernest Ma proposed that light neutrino masses could originate from a neutrinophilic Higgs doublet $\Phi_\nu$ with MeV scale VEV $v_\nu$\cite{Ma:2000cc}. In this way, the heavy neutral leptons are naturally below the TeV scale, which leads to various observable signatures \cite{Ma:2001mr,Haba:2011nb,Haba:2012ai,Maitra:2014qea,Chakdar:2014ifa,Seto:2015rma,Haba:2020lqv}.

Due to the Majorana nature of heavy neutral leptons in seesaw models, the intrinsic collider signature is from the lepton number violation process as $pp\to W^{\pm *}\to \ell^\pm N\to\ell^\pm\ell^\pm j j$ \cite{delAguila:2008cj,Atre:2009rg,Dev:2013wba,Alva:2014gxa,Banerjee:2015gca,Deppisch:2015qwa,Cai:2017mow}. Meanwhile, if a heavy neutral lepton is lighter than the $W$ boson, it becomes long-lived with a small enough mixing parameter, and then leaves a displaced vertex (DV) signature inside the detectors \cite{Cottin:2018nms,Drewes:2019fou,Liu:2019ayx}. The LHC experiment has now excluded the parameter region with $|V_{\ell N}|^2\gtrsim10^{-6}$ and 2 GeV $<m_N<$ 15 GeV \cite{CMS:2022fut,ATLAS:2022atq}. With a relatively clean SM background, the DV signature has drawn increasing attention \cite{Alimena:2019zri,Feng:2024zfe}. Displaced heavy neutral lepton from Higgs decay \cite{Gago:2015vma,Accomando:2016rpc,Accomando:2017qcs,Deppisch:2018eth,Jana:2018rdf,Liu:2022ugx,Bernal:2023coo}, $Z$ decay \cite{Abada:2018sfh,Blondel:2022qqo,Aleksan:2024hyq}, $W'$ decay \cite{Helo:2013esa,Cottin:2018kmq,Nemevsek:2018bbt,Urquia-Calderon:2023dkf}, and $Z'$ decay \cite{Deppisch:2019kvs,Das:2019fee,Chiang:2019ajm,Padhan:2022fak,Bandyopadhyay:2022mej,Li:2023dbs,Liu:2023klu} are extensively studied. In this paper, we consider the heavy neutral leptons from the neutrinophilic Higgs doublet $\Phi_\nu$ decay.

This model introduces a new Higgs doublet $\Phi_\nu$ with lepton number $L_{\Phi_\nu}=-1$, while the heavy neutral leptons $N$ have zero lepton number. Such charge assignment forbids the Yukawa interaction $\bar{L}\tilde{\Phi} N$, but allows the new term $\overline{L}\widetilde{\Phi}_\nu N$ under a global $U(1)_L$ symmetry. When the heavy neutral leptons $N$ are lighter than the neutrinophilic doublet scalars, they can produced via the new decay channels as ${H^\pm}\rightarrow{\ell^\pm}N,{A/H}\rightarrow \nu_\ell N $~\cite{Guo:2017ybk}. The further decay of heavy neutral lepton generates a DV signature with proper mixing parameter. Due to the doublet nature of neutrinophilic scalars, the cross sections of generation processes as $pp\rightarrow{H^+}{H^-}, {H^\pm}{A/H}$ only depend on the scalar masses. Therefore, the cross section of displaced vertex signature from neutrinophilic scalars decay is not suppressed by the mixing parameter $V_{\ell N}$. This new channel is expected to probe $V_{\ell N}$ down to the natural seesaw value. The decay mode $H^\pm\to \ell^\pm N$ also makes this new channel distinguishable from other DV channels.

This work is organized as follows. In Sec. \ref{SEC:TM}, we briefly review the neutrinophilic two Higgs doublet model ($\nu$2HDM) and relevant constraints. In Sce. \ref{SEC:PD}, we discuss  the decay properties of charged scalar and heavy neutral lepton. In Sec. \ref{SEC:DV}, we study the displaced vertex signature of heavy neutral leptons from neutrinophilic scalars at the 14 TeV LHC and 3 TeV CLIC. The conclusions are presented in Sec.\ref{SEC:CL}.

\section{The Model}\label{SEC:TM}

This model further extends the SM by a neutrinophilic scalar doublet $\Phi_\nu$ and three heavy neutral leptons $N$. To forbid the Yukawa interaction of heavy neutral leptons to SM Higgs doublet $\bar{L}\tilde{\Phi} N$,  a global $U(1)_L$ symmetry is imposed with $L_{\Phi_\nu}=-1$ and $L_N=0$. A small VEV of $\Phi_\nu$ can be naturally induced by a soft $U(1)_L$ breaking term $\mu^2(\Phi^\dag \Phi_\nu+\rm{h.c.})$. In this way, the neutrinophilic doublet $\Phi_\nu$ couples to the heavy neutral leptons via the Yukawa interaction $\bar{L}\tilde{\Phi}_\nu N$, which is responsible for the generation of tiny neutrino masses. The new terms are given by
\begin{equation}
	-\mathcal{L}_N=y\overline{L}\widetilde{\Phi}_\nu{N}+\frac{1}{2} m_{N}\overline{N^c}N+\rm{h.c.}, \label{Eqn:Yukawa}
\end{equation}
with $\widetilde{\Phi}_\nu=i\sigma_2{\Phi}_\nu^*$. Similar to the type I seesaw mechanism, light neutrino masses can be derived as,
\begin{equation}
	m_\nu=-\frac{1}{2}v_\nu^2{y}m_{N}^{-1}y^T.\label{Eqn:mv}
\end{equation}

The mixing parameter between the heavy and light neutrinos is expressed as~\cite{FileviezPerez:2009hdc}
\begin{equation}
	V_{lN}=\frac{yv_\nu}{\sqrt{2}}m_{N}^{-1}=U_{\rm{PMNS}}\hat{m}_\nu^{1/2}Rm_{N}^{-1/2},\label{Eqn:VlN}
\end{equation}
where $U_\text{PMNS}$ is the neutrino mixing matrix, $\hat{m}_\nu={\rm diag}(m_1,m_2,m_3)$ is the diagonalized neutrino mass matrix, and $R$ is a generalized orthogonal matrix. Since the natural seesaw predicted value is usually too small to be detected, we consider $V_{\ell N}$ as free parameters in this paper. Moreover, it is sufficient to consider one heavy neutral lepton mixing with light neutrinos for the collider signature as 
\begin{equation}
	\mathcal{L}\supset -\frac{g}{\sqrt{2}} W_\mu \bar{N} V_{\ell N}^{*} \gamma^\mu P_L \ell  
	-\frac{g}{2\cos\theta_W} Z_\mu \bar{N}  V_{\ell N}^{*}  \gamma^\mu P_L \nu_\ell   - \frac{g m_N}{2 m_W} h \bar{N} V_{\ell N}^{*} P_L \nu_\ell   + \text{h.c.}.
\end{equation}

The two scalar doublets can be denoted as,
\begin{align}
\Phi=\left(
\begin{array}{c}
\phi^+\\
\frac{1}{\sqrt{2}}(v+\phi^{0,r}+i\phi^{0,i})
\end{array}
\right), \quad
\Phi_\nu =
\left(
\begin{array}{c}
\phi_\nu^+\\
\frac{1}{\sqrt{2}}(v_\nu+\phi^{0,r}_\nu+i\phi^{0,i}_\nu)
\end{array}\right).
\end{align}
The scalar potential under the global $U(1)_L$ symmetry is,
\begin{eqnarray}\label{Eqn:V}
V&=-&m_\Phi^2(\Phi^\dag \Phi)+m_{\Phi_\nu}^2(\Phi^\dag_\nu \Phi_\nu)+\frac{1}{2}\lambda_1(\Phi^\dag \Phi)^2+\frac{1}{2}\lambda_2(\Phi^\dag_\nu \Phi_\nu)^2 \\\nonumber 
&&+\lambda_3(\Phi^\dag \Phi)(\Phi^\dag_\nu \Phi_\nu)+\lambda_4(\Phi^\dag \Phi_\nu)(\Phi^\dag_\nu \Phi)-\mu^2(\Phi^\dag \Phi_\nu+\text{h.c.}),
\end{eqnarray}
where the $\mu^2$-term breaks the $U(1)_L$ symmetry explicitly. The boundedness condition of
the potential requires \cite{Gunion:2002zf}
\begin{equation}
	\lambda_1,\lambda_2>0,~\lambda_3+\sqrt{\lambda_1 \lambda_2}>0,~\lambda_3+\lambda_4+\sqrt{\lambda_1 \lambda_2}>0
\end{equation}

Assuming $\mu^2<<m_{\Phi_\nu}^2$, we can obtain the relations of VEVs by deriving the minimization conditions
\begin{equation}
	v\simeq\sqrt{\frac{2m_\Phi^2}{\lambda_1}},\quad
	v_\nu\simeq\frac{\mu^2v}{m_{\Phi_\nu}^2+(\lambda_3+\lambda_4)v^2/2}.
\end{equation}
For an electroweak scale $m_{\Phi_\nu}$, $v_\nu\sim\mathcal{O}({\rm MeV})$ is obtained with $\mu^2\sim1~{\rm GeV}^2$. The smallness of $\mu^2$ is protected by the soft broken $U(1)_L$ symmetry. The VEV hierarchy $v_\nu\ll v$ indicates that mixings between the two Higgs doublets are heavily suppressed by $v_\nu/v$. Masses of the physical Higgs bosons are
\begin{eqnarray}
	m_h^2 &\simeq& \lambda_1 v^2, \label{Eqn:mh}\\  
	m_H^2 &\simeq& m_{\Phi_\nu}^2 + \frac{1}{2}(\lambda_3+\lambda_4)v^2, \label{Eqn:mH}\\  
	m_A^2 &\simeq& m_{\Phi_\nu}^2 + \frac{1}{2}(\lambda_3+\lambda_4)v^2, \\  
	m_{H^\pm}^2 &\simeq& m_{\Phi_\nu}^2 +\frac{1}{2}\lambda_3 v^2,
\end{eqnarray}
where the terms of $\mathcal{O}(v_\nu^2)$ and $\mathcal{O}(\mu^2)$ are neglected.
For simplicity, a degenerate mass spectrum of neutrinophilic scalars $m_{H^\pm}=m_H=m_A$ is assumed in the following discussion, which is realized with vanishing $\lambda_4$.

The neutrinophilic doublet could induce observable lepton flavor violation processes \cite{Bertuzzo:2015ada}. The most stringent constraint comes from the radiative decay $\mu\to e\gamma$ with the experimental limit BR$(\mu\to e\gamma)<4.2\times10^{-13}$ \cite{MEG:2016leq}. The predicted branching ratio is \cite{Guo:2017ybk}
\begin{equation}
	\label{BRLFV}
	\text{BR}(\mu\to e\gamma)\approx\frac{3\alpha}{16\pi G_F^2} \frac{m_{N}^2
		|\widetilde{m}_{\mu e}|^2}{m_{H^+}^4 v_\nu^4} \left|F\left(\frac{m_N^2}{m_{H^+}^2}\right)\right|^2,
\end{equation}
with $\widetilde{m}=U_{\text{PMNS}}\hat{m}_\nu U^\dag_{\text{PMNS}}$, and the loop function $F(x)$ is
\begin{equation}
	F(x)=\frac{1}{6(1-x)^4}\left(1-6x+3x^2+2x^3-6x^2\ln x\right).
\end{equation}
To satisfy the current experimental limit, $m_{H^\pm}\cdot v_\nu\gtrsim 600\text{GeV}\cdot\text{MeV}$ is required for electroweak scale $m_N$. In this paper, we fix $v_\nu=10$~MeV, which results in the corresponding Yukawa coupling $y\sim\sqrt{2 m_\nu m_N}/v_\nu\sim 10^{-2}$ for $m_\nu\sim0.05$ eV and $m_N\sim100$ GeV.

Searches for the neutrinophilic doublet mainly focus on the charged scalar $H^\pm$. When the heavy neutral lepton $N$ is heavier than the charged scalar $H^\pm$, the leptonic decay $H^\pm\to\ell^\pm \nu$ is the dominant channel \cite{Haba:2011nb}, which requires $m_{H^\pm}\gtrsim700$ GeV to satisfy the LHC limit \cite{ATLAS:2019lff,CMS:2020bfa}. On the other hand, the charged scalar decays to leptons and heavy neutral lepton as $H^\pm\to \ell^\pm N$ when $m_N<m_{H^+}$. Further decays of $N$ lead to several lepton number violation signatures \cite{Guo:2017ybk}. Since there are no experiments searching for such signatures, we take the LEP bonds on charged scalar mass $m_{H^\pm}>80\rm{GeV}$ \cite{ALEPH:2013htx}. In this study, we consider the light HNL scenario $m_N<m_{H^+}$. We also assume $m_{H^+}\geq200$ GeV to satisfy the direct search limit.

\section{Decay Properties}\label{SEC:PD}

\begin{figure}
	\begin{center}
		\includegraphics[width=0.45\linewidth]{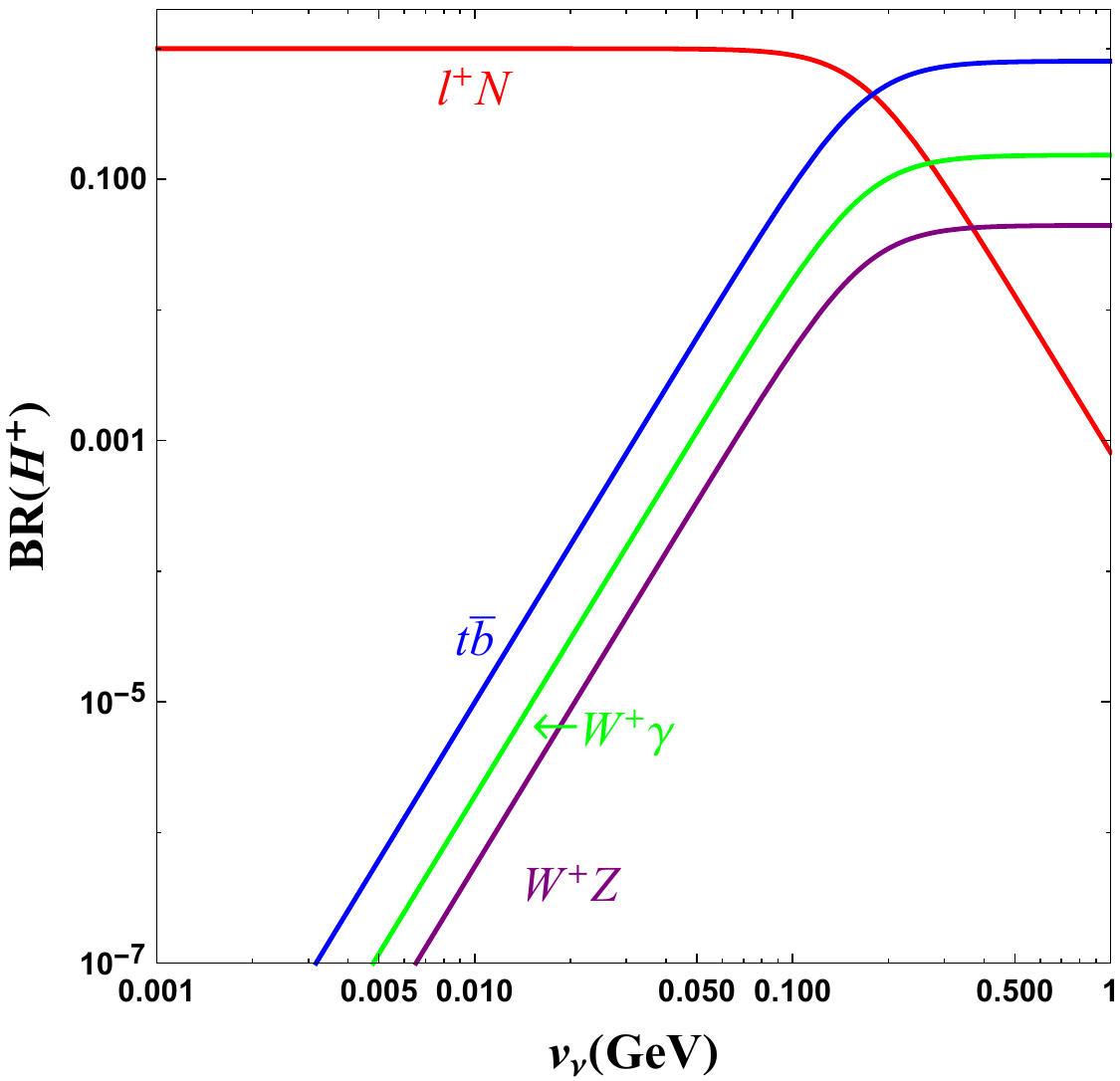}
		\includegraphics[width=0.47\linewidth]{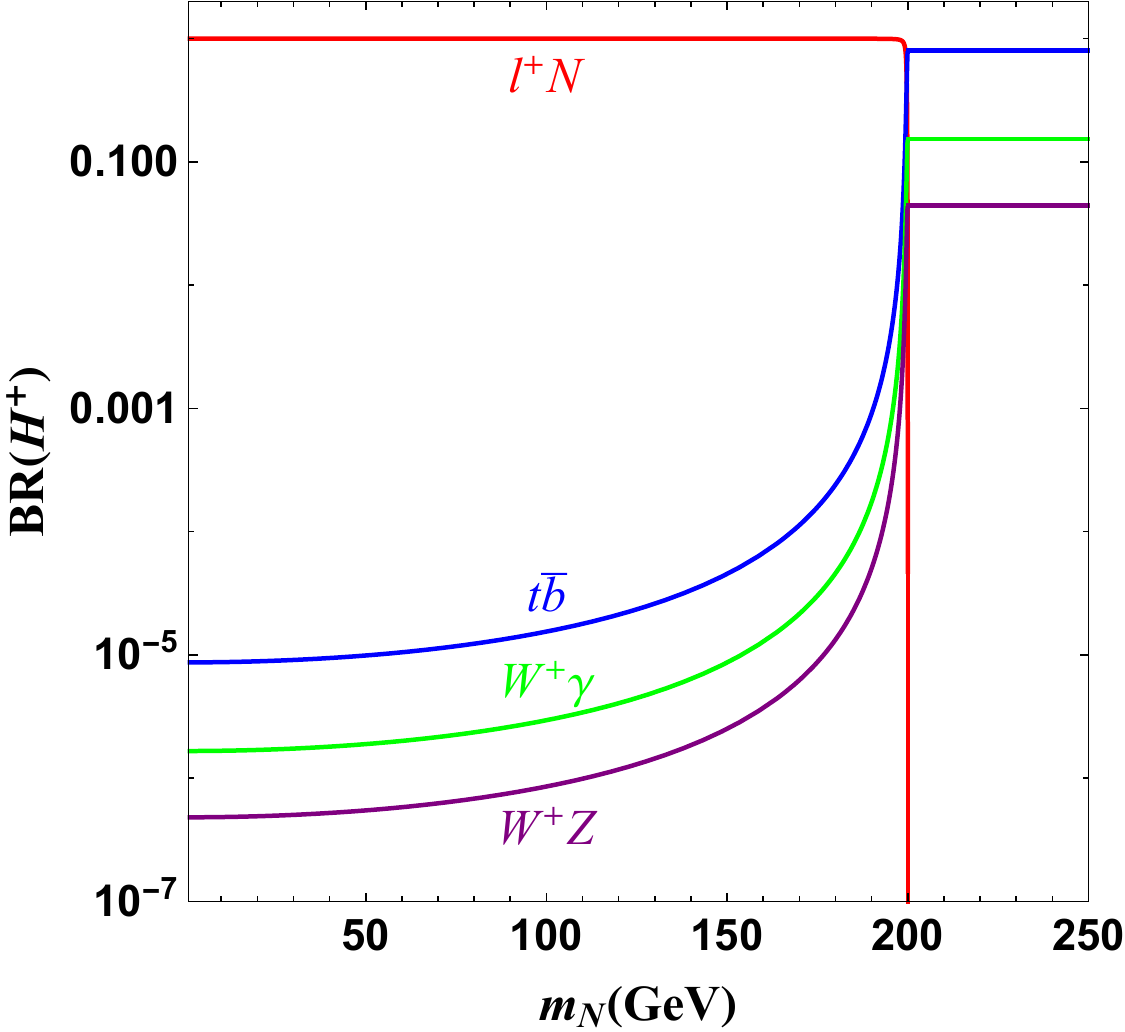}
	\end{center}
	\caption{Branch ratio of charged scalar $H^+$ with $m_{H^+}=200~\rm{GeV}$. In the left panel, we assume $m_N=100$ GeV. In the right panel, we have fixed $v_\nu=10$ MeV.}
	\label{hc}
\end{figure}

The decay properties of the neutrinophilic scalars $H^{\pm}$, $A$, $H$ and heavy neutral lepton $N$ have been discussed in Refs.~\cite{Ma:2000cc,Haba:2011nb}. In our consideration of $m_{H^+}>m_{N}$, the neutrinophilic charged scalar can decay into leptons and heavy neutral lepton via the neutrino Yukawa interaction as,
 \begin{eqnarray}
\Gamma(H^+\rightarrow{l^+}N)=\frac{|y|^2m_{H^+}}{16\pi}\left(1-\frac{m_{N}^2}{m_{H^+}^2}\right)^2,
\end{eqnarray}
where the Yukawa coupling $y$ is typically at the order of $\mathcal{O}(10^{-2})$.
In addition, $H^+$ can also decay into $t\bar{b},W^+\gamma,W^+Z$ via mixing with SM Higgs doublet, which is suppressed by the mixing factor $v_\nu/v\sim 10^{-4}$ as $v_\nu=10$ MeV in the following studies. In Figure~\ref{hc}, we show the branching ratio of charged scalar $H^+$. The leptonic decay $H^+\to l^+ N$ is the dominant channel when $v_\nu\lesssim0.1$ GeV. For $v_\nu=10$ MeV, the $H^+\to l^+ N$ is the dominant decay mode once it is kinematically allowed, meanwhile, the neutral scalars decay into neutrinos as $H/A\to \nu N$. 

\begin{figure}
	\begin{center}
		\includegraphics[width=0.45\linewidth]{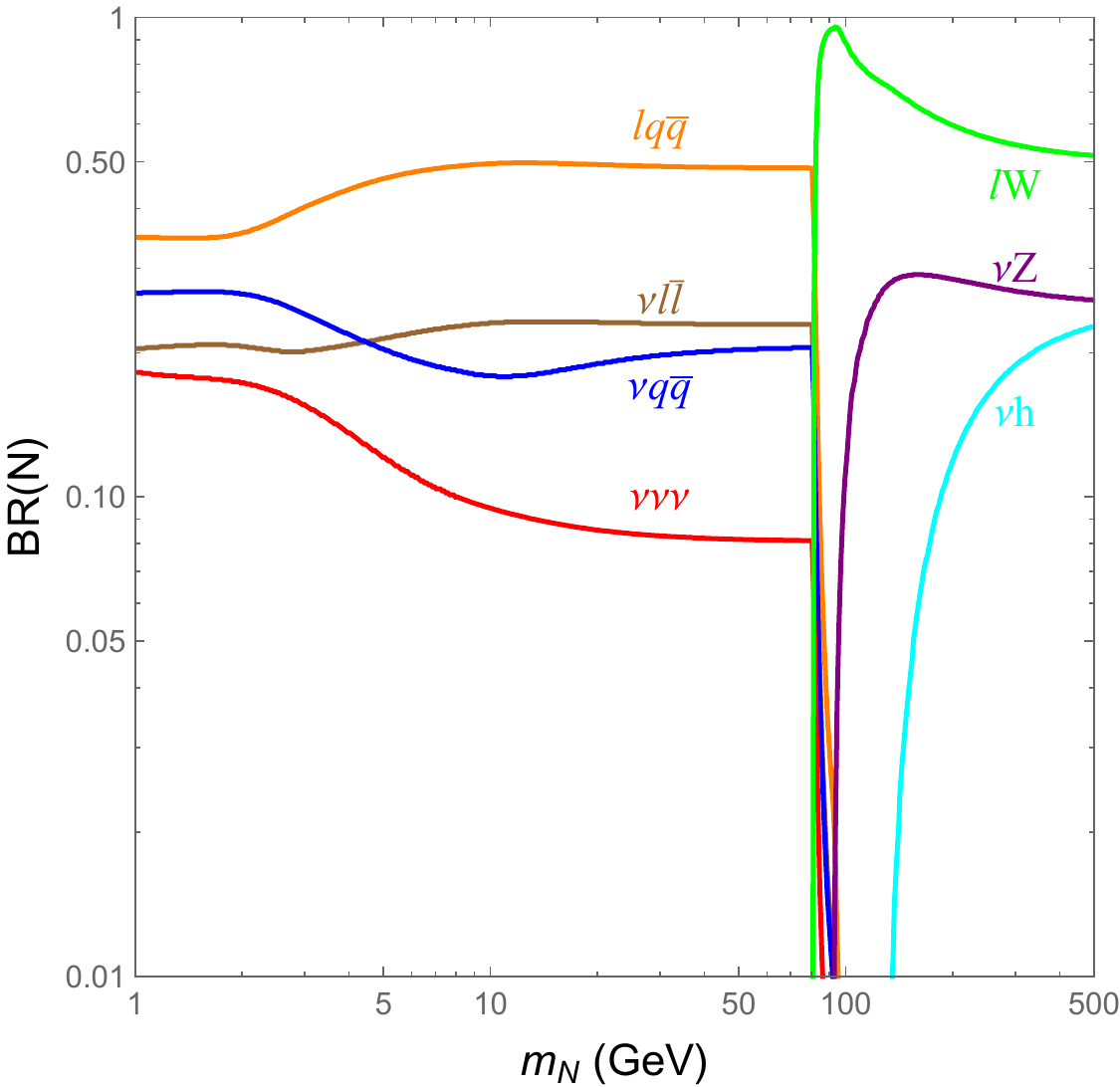}
		\includegraphics[width=0.45\linewidth]{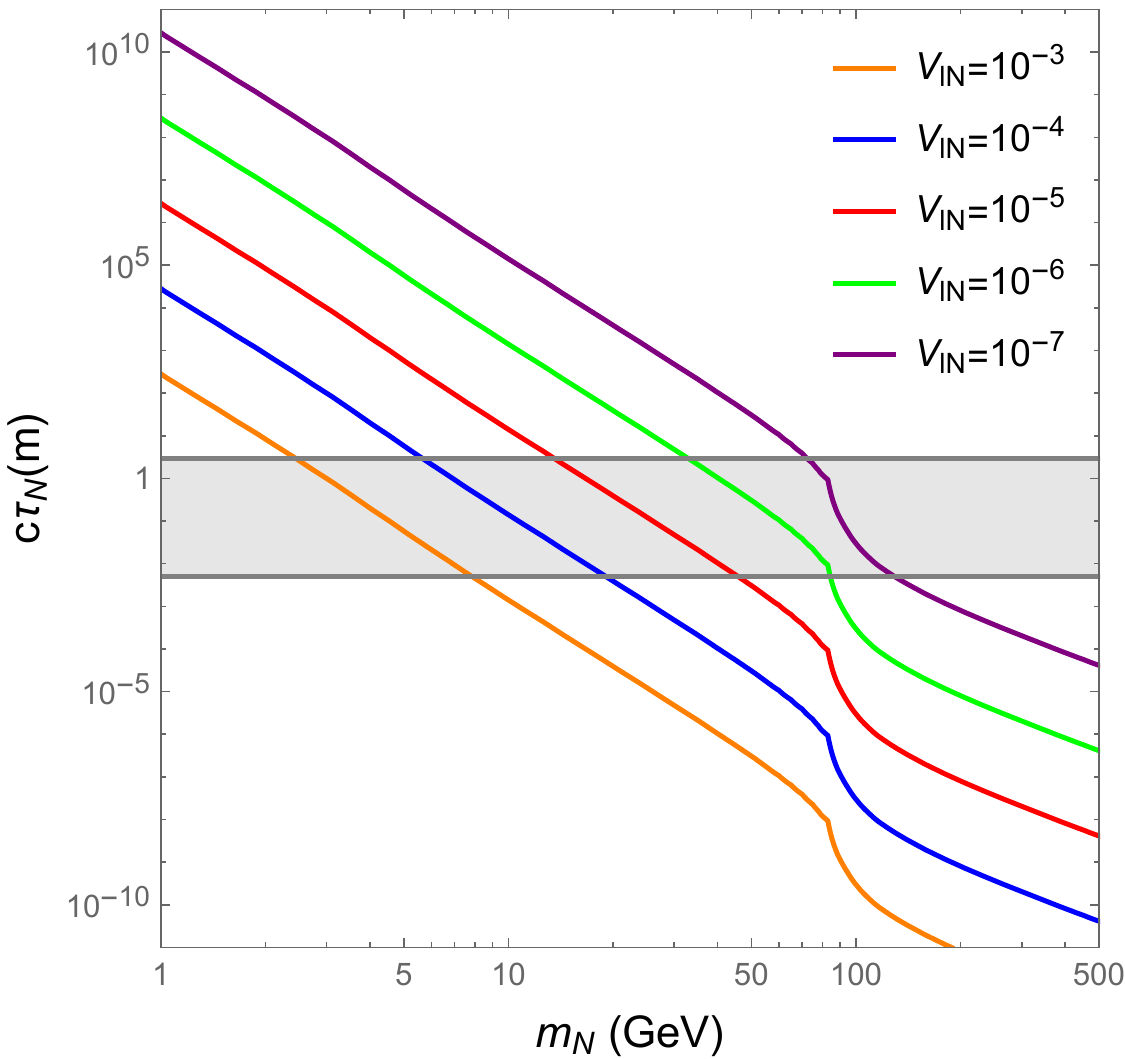}
	\end{center}
	\caption{Decay branch ratio of heavy neutral lepton $N$ (left) and the theoretical decay length (right). The gray bond in the right panel is the collider sensitive region.}
	\label{n1}
\end{figure}

In the scenario with $m_N<m_{H^+}$, the decay widths of heavy neutral lepton $N$ are determined by its mass $m_N$ and the mixing parameter $V_{lN}$. When $m_N<m_W$, the heavy neutral lepton $N$ decays into three fermions via off-shell $W$ and $Z$. For heavier $m_N$, the two-body decays $N\to l^\pm W^\mp,\nu Z,\nu h$ become the dominant decay channels. See Ref.~\cite{Atre:2009rg} for the explicit decay width of each channel. In Figure~\ref{n1}, we show the branching ratio of heavy neutral lepton and the decay length $c\tau_N$. To match the collider sensitive region, the heavier the $M_N$ is, the smaller the $V_{\ell N}$ is required. Different from the canonical type-I seesaw, the production of heavy neutral lepton is via decays of neutrinophilic scalars, which is not suppressed by the mixing $V_{\ell N}$. Therefore, we expect to probe larger parameter space, i.e., smaller $V_{\ell N}$ and heavier $m_N$.

\section{Displaced Vertex Signatures}\label{SEC:DV}

\begin{figure}
	\begin{center}
		\includegraphics[width=0.6\linewidth]{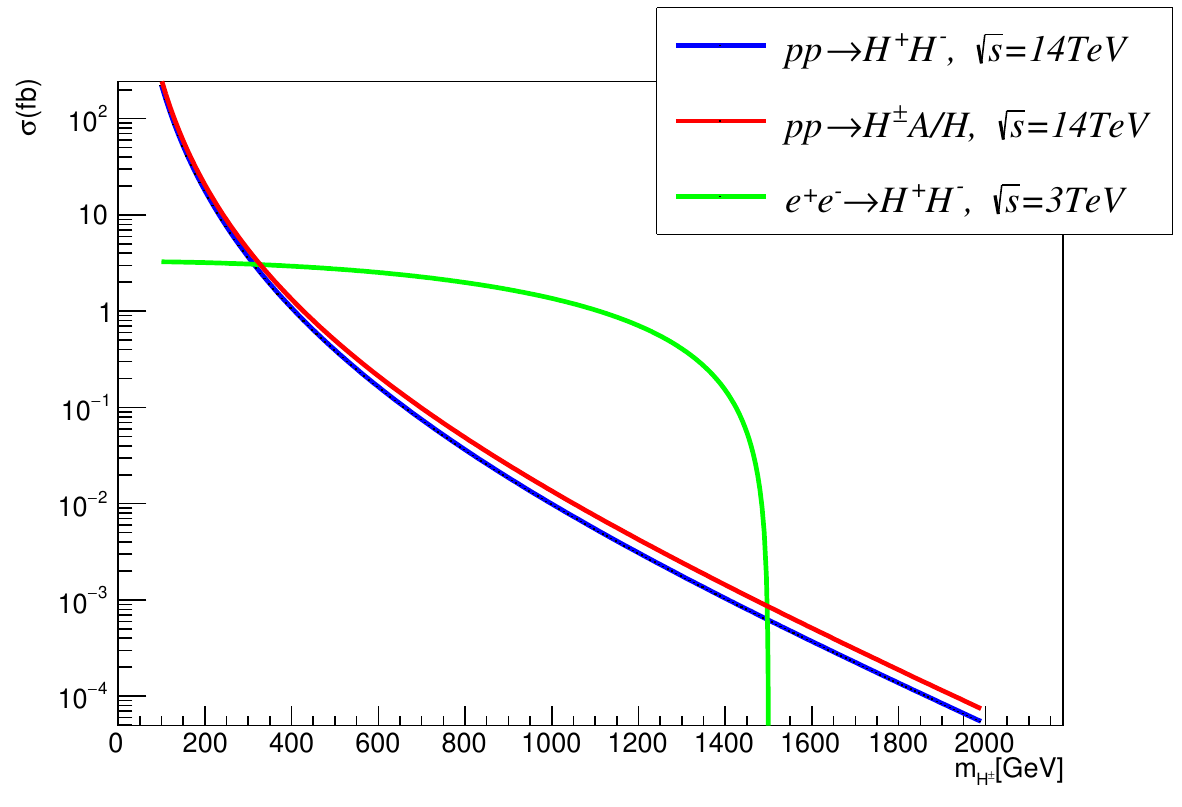}
	\end{center}
	\caption{Cross section of neutrinophilic scalars at 14 TeV LHC and 3 TeV CLIC.}
	\label{cs}
\end{figure}

The long-lived heavy neutral leptons travel a macroscopic distance within the detector before decaying, which leads to the so-called DV signature. Due to negligible SM backgrounds, this signature is a unique way to probe GeV scale $m_N$ with proper $V_{\ell N}$ at colliders and beam-dump experiments. In this paper, we study the DV signature of heavy neutral leptons from neutrinophilic scalars at the 14~TeV LHC and 3 TeV CLIC. In Figure \ref{cs}, we show the cross section of neutrinophilic scalars at 14 TeV LHC and 3 TeV CLIC. The corresponding production processes are shown in Figure \ref{pp} and Figure \ref{ee}. For a naive estimation, LHC and CLIC could probe charged scalar up to TeV order. 

To simulate the displaced vertex signature, we use {\bf FeynRules2.3} \cite{Alloul:2013bka}  to generate the Universal FeynRules Output (UFO) \cite{Degrande:2011ua} file of the $\nu\rm{2HDM}$ model. Then events are generated by {\bf Madgraph5\_aMC@NLO} \cite{Alwall:2014hca}, and {\bf Pythia8}\cite{Sjostrand:2007gs} is used to do parton showering and hadronization. The detector simulation is performed by {\bf Delphes3}\cite{deFavereau:2013fsa} with the corresponding cards for LHC and CLIC respectively.

\subsection{Signature at LHC}

\begin{figure}
	\begin{center}
		\includegraphics[width=0.45\linewidth]{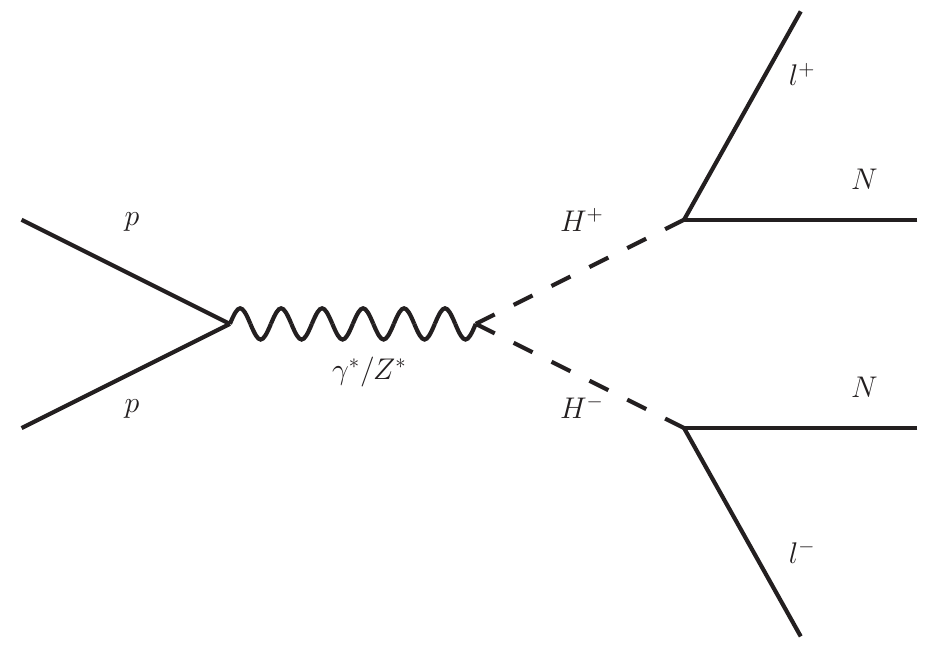}
		\includegraphics[width=0.45\linewidth]{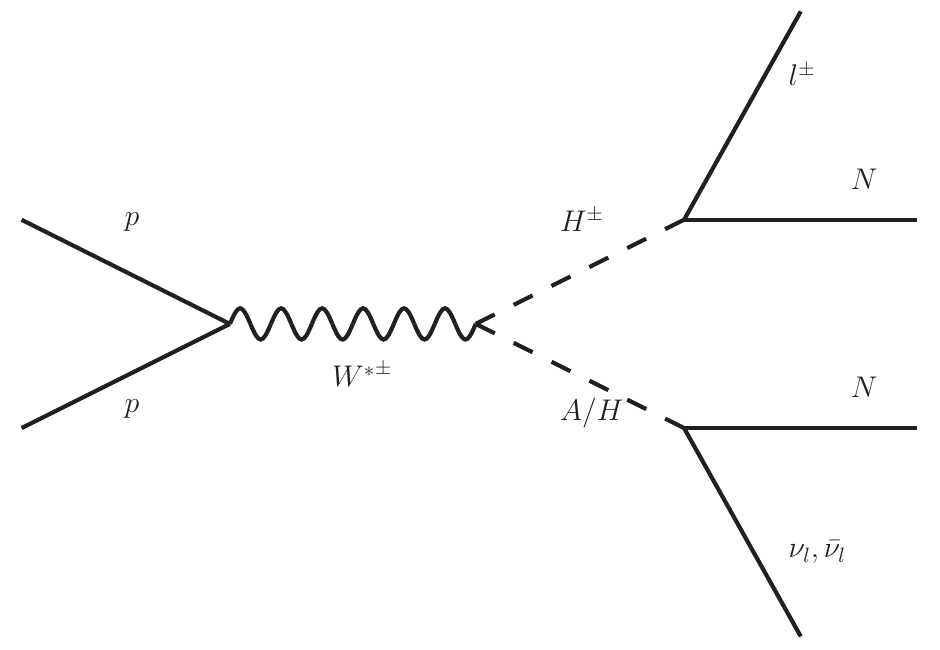}
	\end{center}
	\caption{Feynman diagrams for the processes of $pp\rightarrow{H^+}{H^-}\rightarrow{\ell^+}N \ell^-N$ in the left panel, and $pp\rightarrow{H^\pm}{A/H}\rightarrow{\ell^\pm}N \nu_\ell N$ in the right panel.
	}
	\label{pp}
\end{figure}

In this section, we aim to search for the DV signature of heavy neutral leptons at 14 TeV LHC. It should be noted that the Yukawa matrix $y$ and mixing pattern $V_{\ell N}$ are sophisticated for the theoretical predictions of the $\nu\rm{2HDM}$.  Following the spirit of the experimental groups, we focus on a simplified scenario, in which neutrinophilic scalars and heavy neutral lepton $N$ only couple to electron or muon exclusively. Due to much lower tagging efficiency, we do not consider the tau channel in this analysis. At the 14~TeV LHC, the signal process can be written as 
\begin{align}
	pp\rightarrow&{H^+}{H^-}\rightarrow{\ell^+}N \ell^-N ,\label{Eqn:signal01}\\
	pp\rightarrow&{H^\pm}{A/H}\rightarrow{\ell^\pm}N \nu_\ell N,\label{Eqn:signal02}
\end{align}
where $\ell=e$ or $\mu$ exclusively. Corresponding signal processes are shown in Figure~\ref{pp}. The pair production of the charged scalar process  $pp\rightarrow{H^+}{H^-}$ is mediated by the virtual $\gamma^*$ and $Z^*$, and the associated production processes $pp\rightarrow{H^\pm}{A/H}$  are mediated by the virtual $W^{*\pm}$. Due to the doublet nature of neutrinophilic scalars, the production cross sections at 14 TeV LHC only depend on their masses. For $m_{H^+}$ less than 1000~GeV, the cross sections are larger than 0.01 fb, which are hopefully to be detected \cite{Guo:2017ybk}. Decay of charged scalar $H^\pm$ leads to a prompt charged lepton. In order to trigger the DV signal, we require at least one prompt charged lepton in the  final states. Therefore, we do not include the contribution of process $pp\to HA\to \nu_\ell N \nu_\ell N$ in the following analysis.

\begin{figure}
	\begin{center}
		\includegraphics[width=0.45\linewidth]{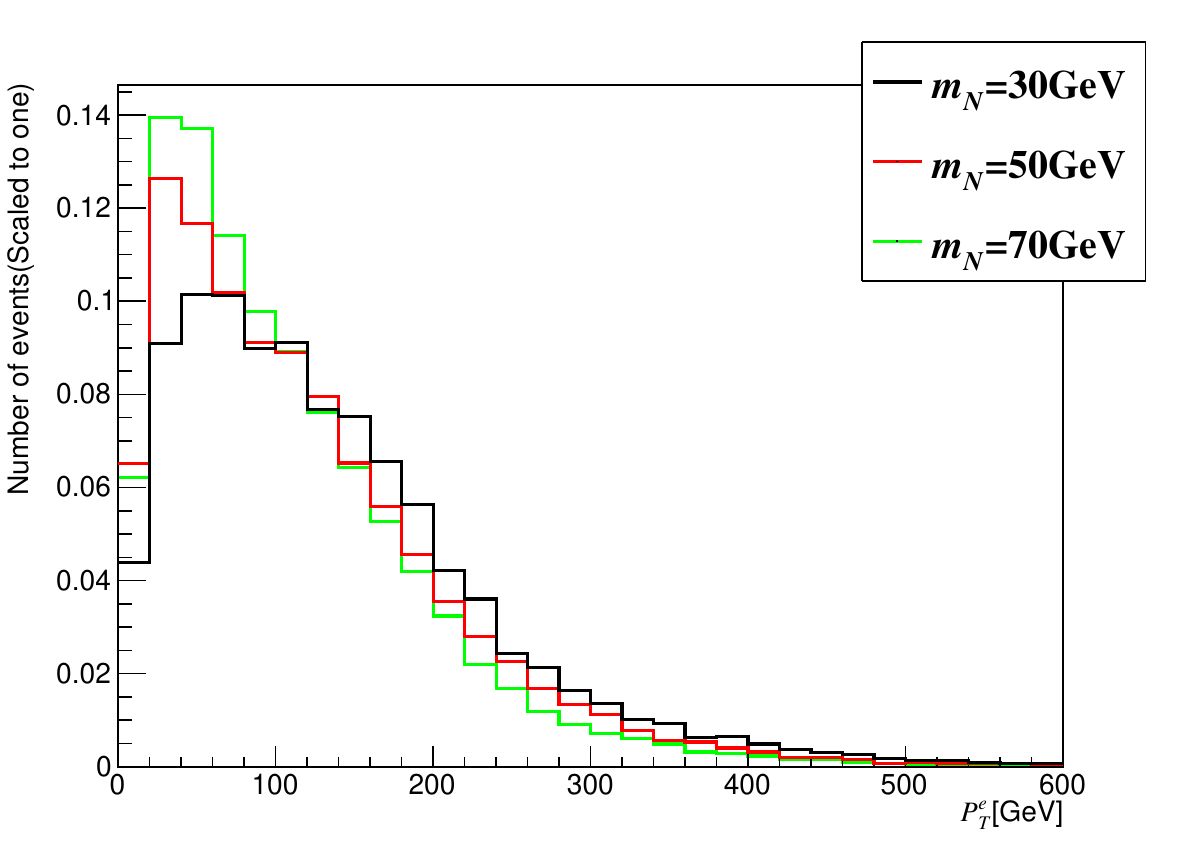}
		\includegraphics[width=0.45\linewidth]{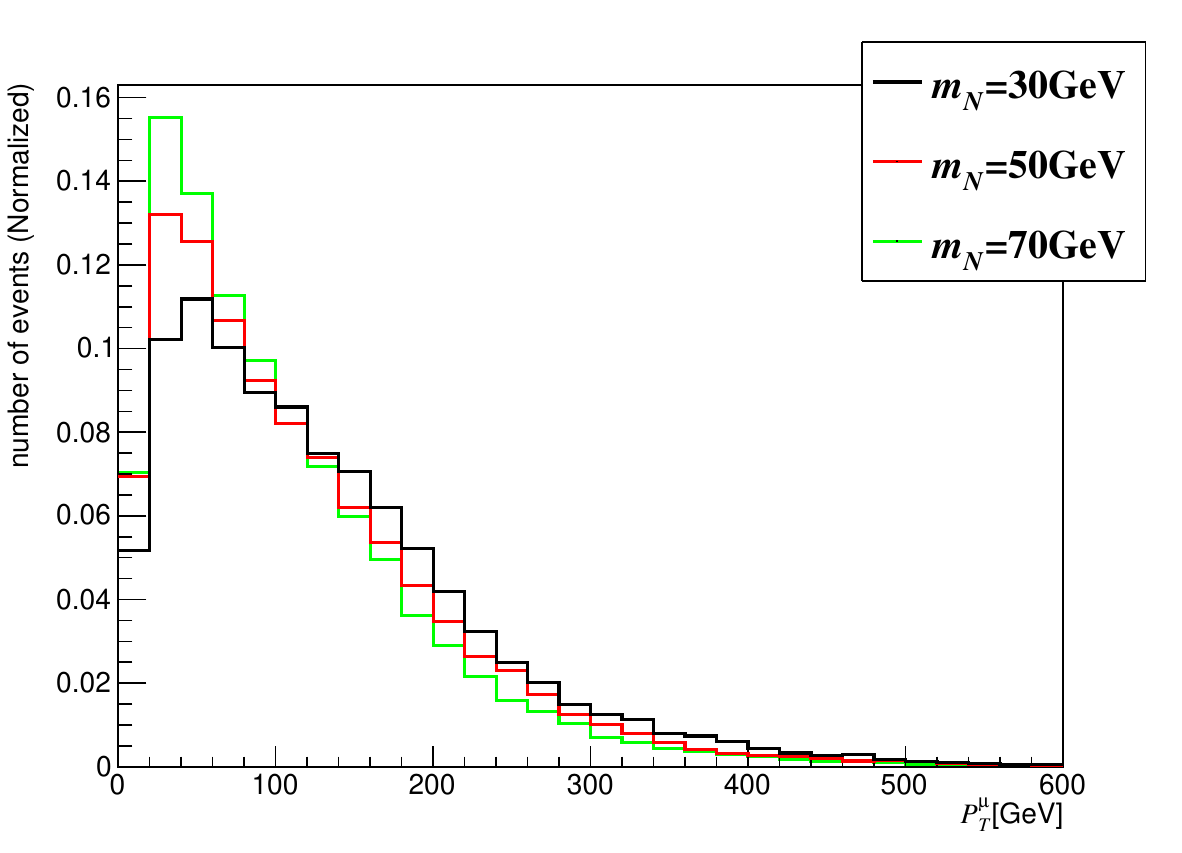}
		\includegraphics[width=0.45\linewidth]{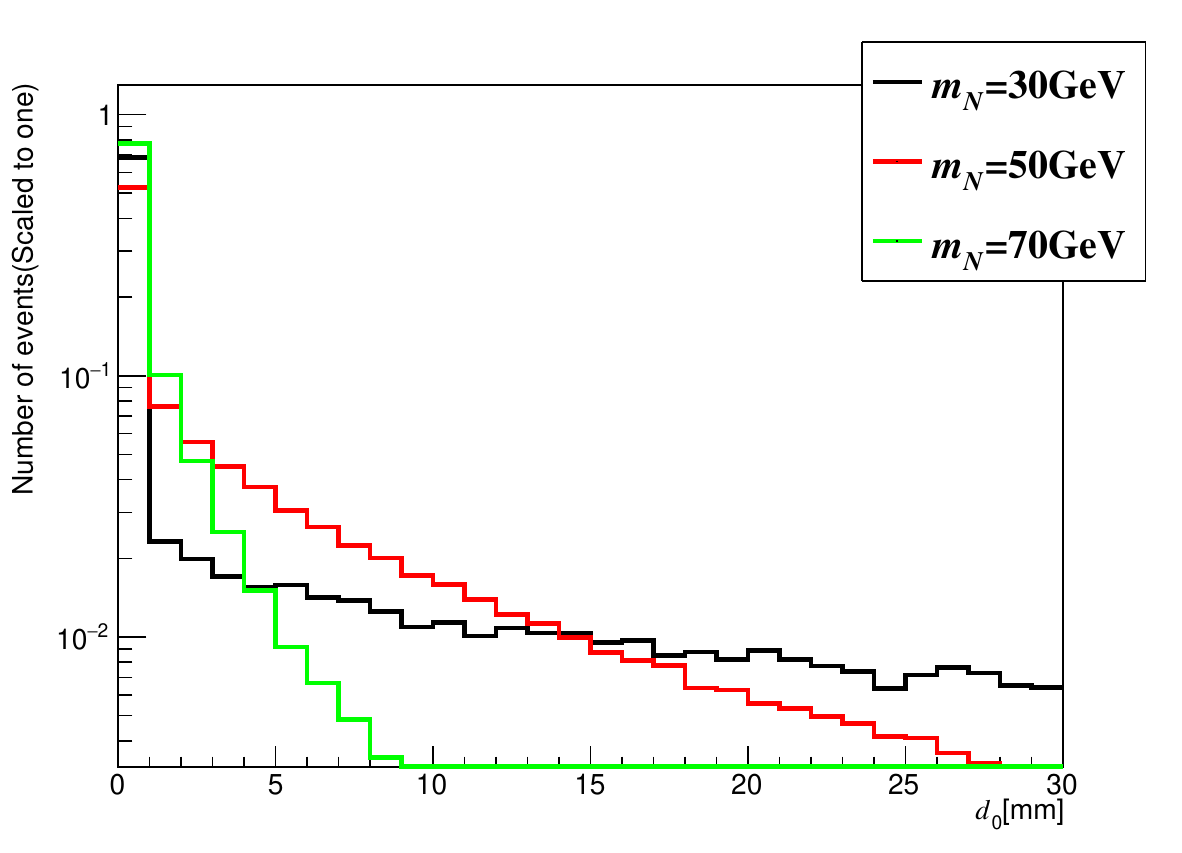}
		\includegraphics[width=0.45\linewidth]{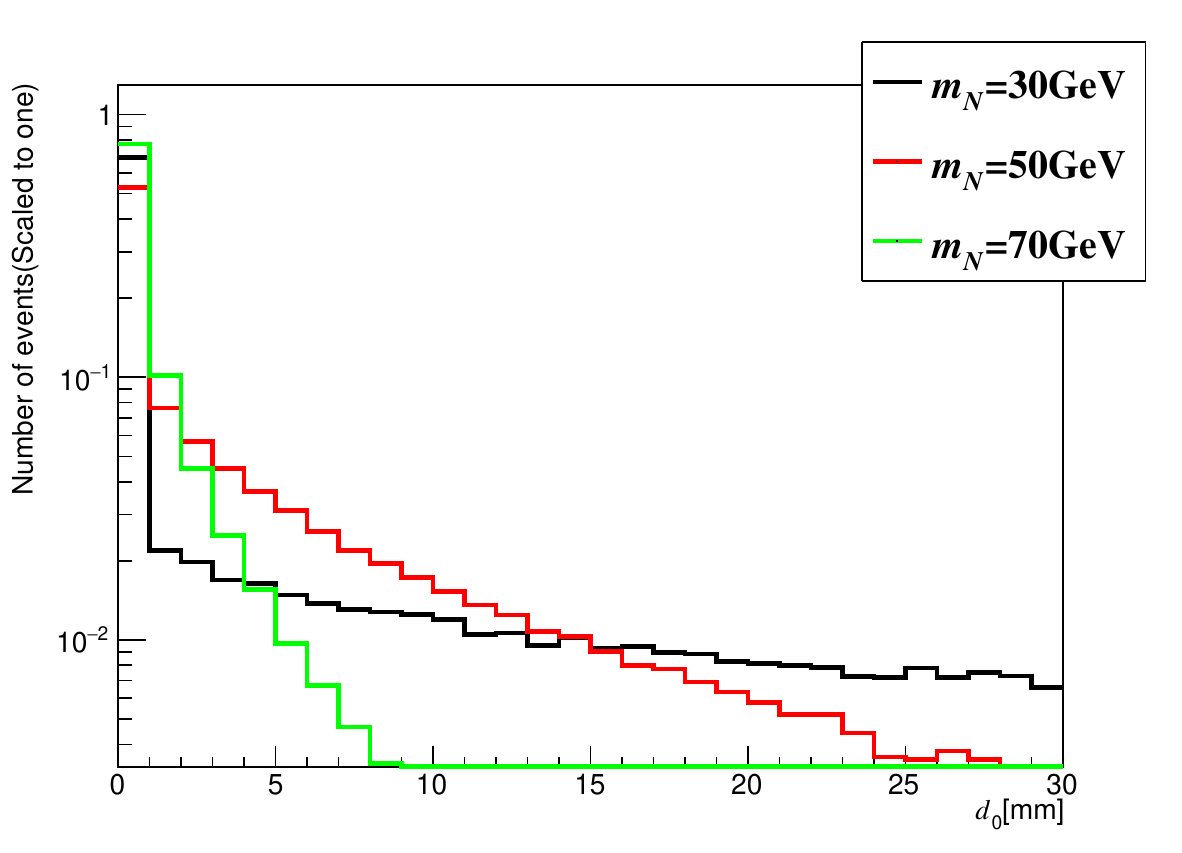}
		\includegraphics[width=0.45\linewidth]{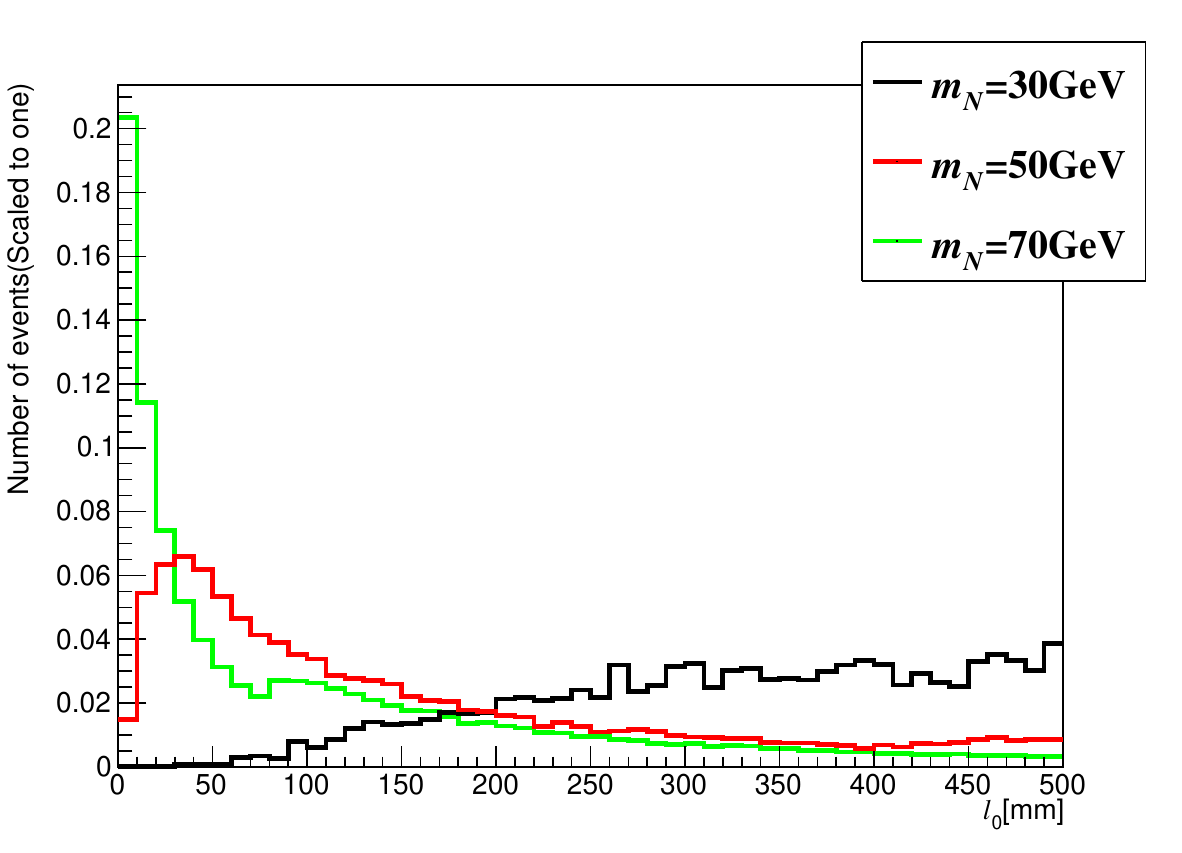}
		\includegraphics[width=0.45\linewidth]{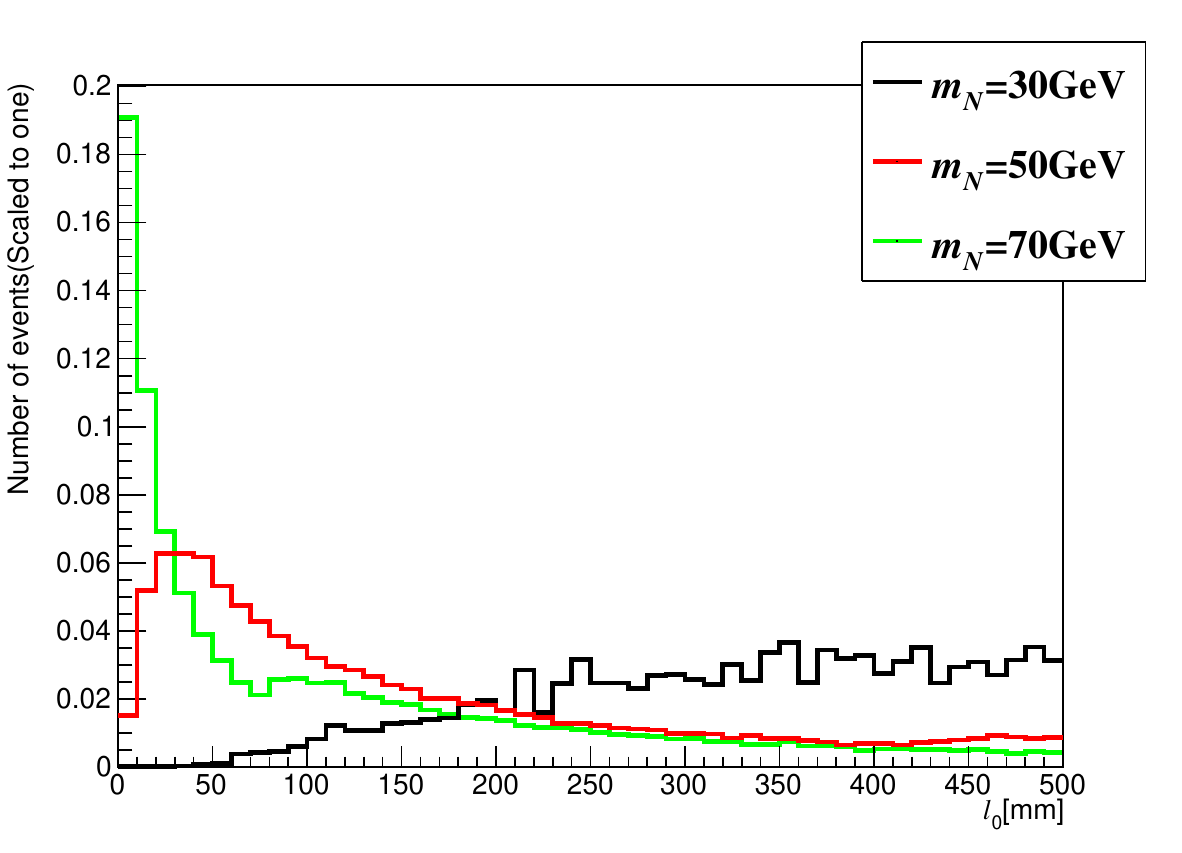}
		\includegraphics[width=0.45\linewidth]{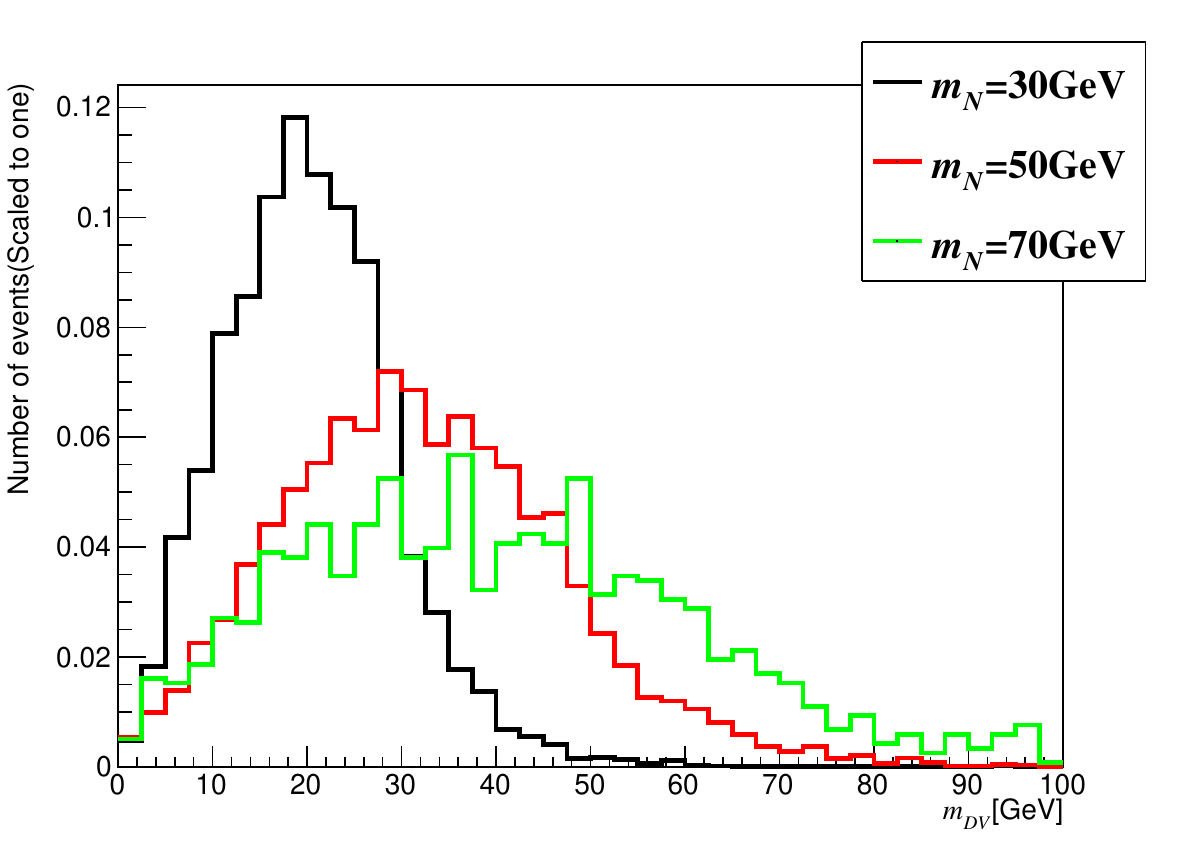}
		\includegraphics[width=0.45\linewidth]{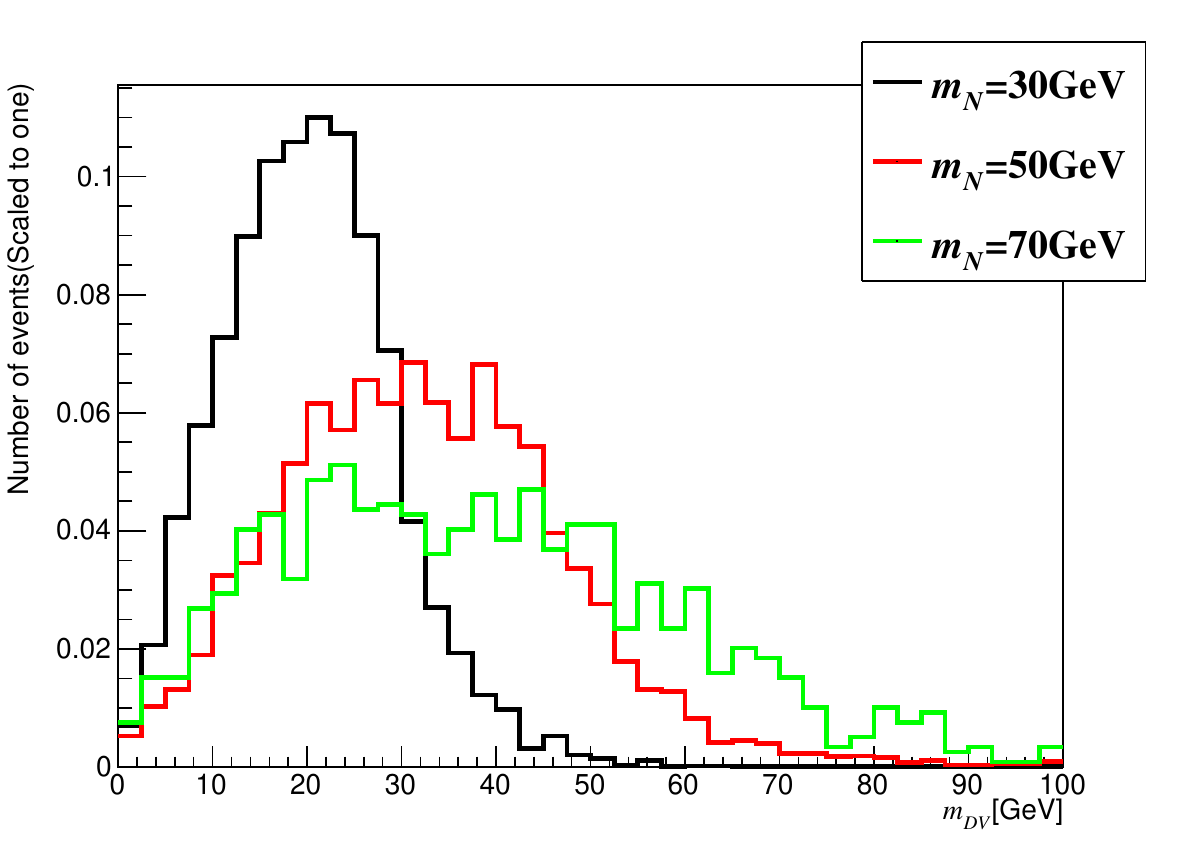}
	\end{center}
	\caption{Distributions of  lepton transverse momentum $P_T^{\ell}$, transverse impact parameter $d_0$, displacement for tracks $l_0$, reconstructed DV mass $m_{DV}$. The left (right) panels are for the electron (muon) channel.  }
	\label{track}
\end{figure}

The further decays of HNL generate the DV signature, which depends on its mass $m_N$ and mixing parameter $V_{\ell N}$ as shown in Figure~\ref{n1}. Theoretically speaking, there are two HNLs in the decay of neutrinophilic scalars, which could lead to two DVs signature. With higher tagging efficiency, the one DV signature would have a wider sensitive region than the two DVs signature in principle. However, the one DV signature corresponds to the inclusive process $pp\to \ell^\pm N+X$ with $X$ denoting the undetected particles, which has an additional contribution from $pp\to W^\pm\to \ell^\pm N$. The latter process could become the dominant channel for relatively large mixing parameter $V_{\ell N}$, which then makes it hard to distinguish various seesaw models with HNL. This ambiguity can be solved when $V_{\ell N}$ is small enough to make $H^\pm\to\ell^\pm N$ the dominant contribution, especially for the region as $m_N>m_W$. On the other hand, the two DVs signature with prompt charged lepton only arises in the $\nu\rm{2HDM}$, but the reconstruction processes for two DVs signal will lost most of the events.

The SM backgrounds of DV signature mainly stem from long-lived heavy flavor hadrons, e.g., $B^0$ meson, which can be effectively excluded by cuts on invariant mass and displacement \cite{Drewes:2019fou}. The backgrounds may also be from the random crossing of tracks that compose a fake DV, from nuclear interactions with the detector material, or from fake photons, cosmic rays, and beam-halo muons, which are hard to simulate and estimate. The full SM background analysis is beyond the scope of this work, but is safe to consider it as one for an optimistic estimation \cite{Drewes:2019fou}. In searching for one DV signature, we aim to figure out the contribution of neutrinophilic scalars, so the inevitable $pp\to W^\pm\to \ell^\pm N$ process is treated as an irreducible background. Since this new physics background is only sensitive within the region of $|V_{\ell N}|^2\gtrsim10^{-9}$ and $m_N\lesssim30$ GeV at LHC, we do not seek further cuts to suppress it. 

Distributions of some related parameters for selection cuts are shown in Figure \ref{track} . Three benchmark points are selected as $m_{H^+}=200$ GeV while $m_{N}=$30 GeV, 50 GeV, and 70 GeV with $|V_{\ell N}|^2=10^{-10}(\ell = e~\rm{or}~\mu)$ respectively. With a relatively clear environment for DVs search at LHC, here we do not include any SM background as a reference. Meanwhile, the cross section of new physics background  $pp\to W^\pm\to \ell^\pm N$ is less than $10^{-3}$ fb with $|V_{\ell N}|^2=10^{-10}$, thus it is also negligible.

After the simulations, we first select events carrying at least one electron (muon) with transverse momentum greater than 20 GeV and pseudo-rapidity of $|\eta_\ell|<2.5$, which are the trigger cuts of the DV signature. For a lighter $m_N$, the prompt lepton tends to be more  energetic. Distributions of $P_T^e$ are similar to those of $P_T^{\mu}$ for the same value of $m_N$, even though the multiplicity of muons is slightly greater than that of electrons due to the different acceptance at the LHC detector.

Then we select tracks by using a softer momentum cut $P^{trk}>5~\rm{GeV}$ to escape the magnetic field, as well as a large transverse impact parameter of $d_0>2~\rm{mm}$. The parameter $d_0$ is defined as
\begin{equation}
d_0=|x^{trk}P_y^{trk}-y^{trk}P_x^{trk}|/P_T^{trk},
\end{equation}
where $x^{trk},y^{trk}$ are track positions in the transverse plane from the primary interaction vertex, $P_x^{trk},P_y^{trk}$ are the $x$- and $y$-components of the track momentum, and $P_T^{trk}=\sqrt{(P_x^{trk})^2+(P_y^{trk})^2}$ is the transverse momentum of a track. With the fixed value of $|V_{\ell N}|^2=10^{-10}$, the maximum value of $d_0$ decreases as $m_N$ increases. In this sense, the case with $m_N=30$ GeV is more promising for DV signature than those with heavier $m_N$. 

\begin{table}
	\begin{center}
		\begin{tabular}{|c | c | c ||c | c | } 
			\hline
			\hline
			& \multicolumn{4}{|c|}{\textbf{Cut-Flow}} \\
			\hline
			\textbf{Trigger on Lepton} & \multicolumn{4}{|c|}{$N_\ell\geq1,$ $P_T^\ell>20~\rm{GeV},$ $|\eta_\ell|<2.5$}  \\
			\hline
			\textbf{Tracks}  & \multicolumn{4}{|c|}{$P^{trk}>5~\rm{GeV},$ $d_0>2~\rm{mm}$ } \\
			\hline
			\textbf{Displaced Vertex} &  \multicolumn{4}{|c|}{$N_{\ell}\geq1,$ $N_{trk}\geq2,$  $\Delta{x}<1{\rm{mm}},\Delta{y}<1{\rm{mm}},\Delta{z}<1{\rm{mm}}$} \\
			\cline{2-5}
			\textbf{Reconstruction}& \multicolumn{4}{|c|}{$\Delta{R}>0.1,$ $l_0\in[5,3000]\rm{mm},$ $m_{DV}\geq5\rm{GeV}$} 	 \\
			\hline
			\hline
			Benchmark Point & 1 Displaced Vertex & Significance & 2 Displaced Vertex & Significance \\
			\hline
			$m_N=30$ GeV & 25.56(29.34) & 1254(1352) & 6.19(6.94)  & 573(610) \\
			\hline
			$m_N=50$ GeV & 26.66(30.16) & 1283(1373) & 6.06(6.93)  & 566(609) \\
			\hline
			$m_N=70$ GeV & 6.11(6.90) & 569(608)  & 0.25(0.25)  & 93(93) \\
			\hline
		\end{tabular}
	\end{center}
	\caption{Cut flow and the final cross sections for DV signature at LHC. The cross sections are in the unit of fb. The results of muon mixing are in brackets. Significance is calculated with an integrated luminosity of 3 ab$^{-1}$.
		\label{Tab:cut}}
\end{table}

Considering that the dominant decay products of HNL always contain an electron (muon) as shown in Figure~\ref{n1}, we require that the  reconstructed displaced vertex has at least one electron (muon) track. In this work, the long-lived HNLs are considered to decay before the muon chamber. Since tracks from the same HNL are expected to share the same origin point, we reconstruct the displaced vertex by requiring that at least two charged tracks satisfy $\Delta{x}<1{\rm{mm}},\Delta{y}<1{\rm{mm}},\Delta{z}<1{\rm{mm}}$, and tracks are isolated with the condition $\Delta{R}=\sqrt{{\Delta\eta}^2+{\Delta\phi}^2}>0.1$. In order to reduce backgrounds from long-lived SM hadrons,  the reconstructed displaced vertex is required to satisfy a displacement cut of $5~{\rm{mm}}<l_0<3000~\rm{mm}$ and an invariant mass cut of $m_{DV}\geq5~\rm{GeV}$. Since the DV signature originates from the HNL, $m_{DV}\lesssim m_N$ is expected. Here, the upper bound on $m_{DV}$ is not applied so as to extract the mass of HNL.   We summarize all of the selection cuts in Table.\ref{Tab:cut}.

 Assuming Poisson distribution, the significance for $n$ observed events is calculated as \cite{Cowan:2010js}
\begin{equation}
	S(n|b)=\sqrt{-2{\rm{ln}}\frac{P(n|b)}{P(n|n)}},
\end{equation}
where $P(n|b)={b^n}e^{-b}/{n!}$ is the Poisson probability, $b$ is the event number of the backgrounds, and $n=b+s$ is the total event number of background and signal. With one background event, we should have at least nine signal events for the discovery of $S(n|b)>5$. Meanwhile, the system uncertainty of signal and background estimation are overlooked.  The significance of benchmark points is also shown in Table \ref{Tab:cut}. For $m_N=30$ GeV and $m_N=50$ GeV, the cross sections after all selection cuts are similar. With 3~ab$^{-1}$ data, the significance could reach about 1300 for one DV pure electron(muon) channel. The significance for two DVs pure electron(muon) channels are smaller than one DV, which could reach about 600. For $m_N=70$~GeV, although it is less promising than the previous two benchmark points, the significance could reach 90 even for the two DV channels.

\begin{figure}
	\begin{center}
		\includegraphics[width=0.45\linewidth]{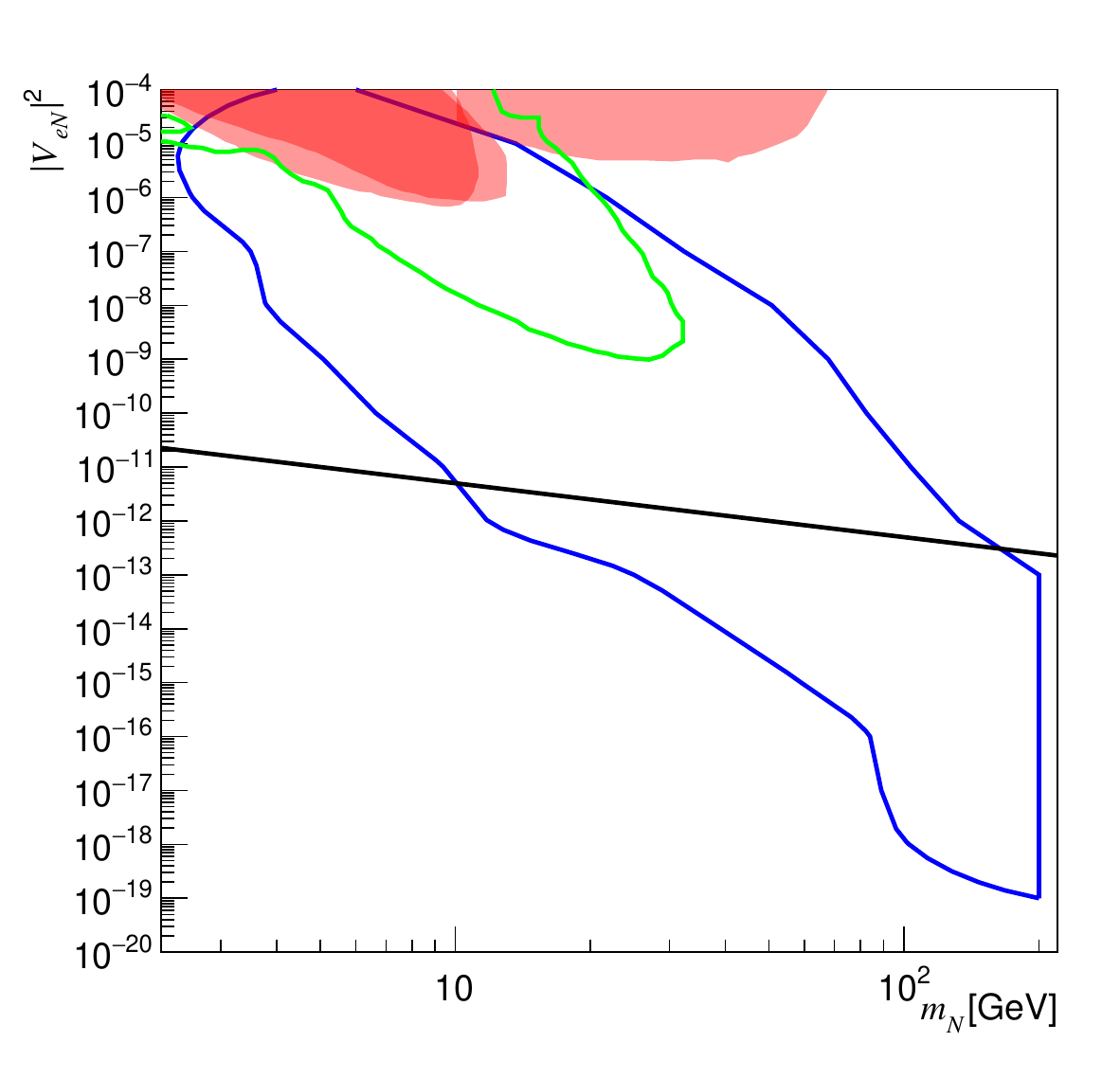}
		\includegraphics[width=0.45\linewidth]{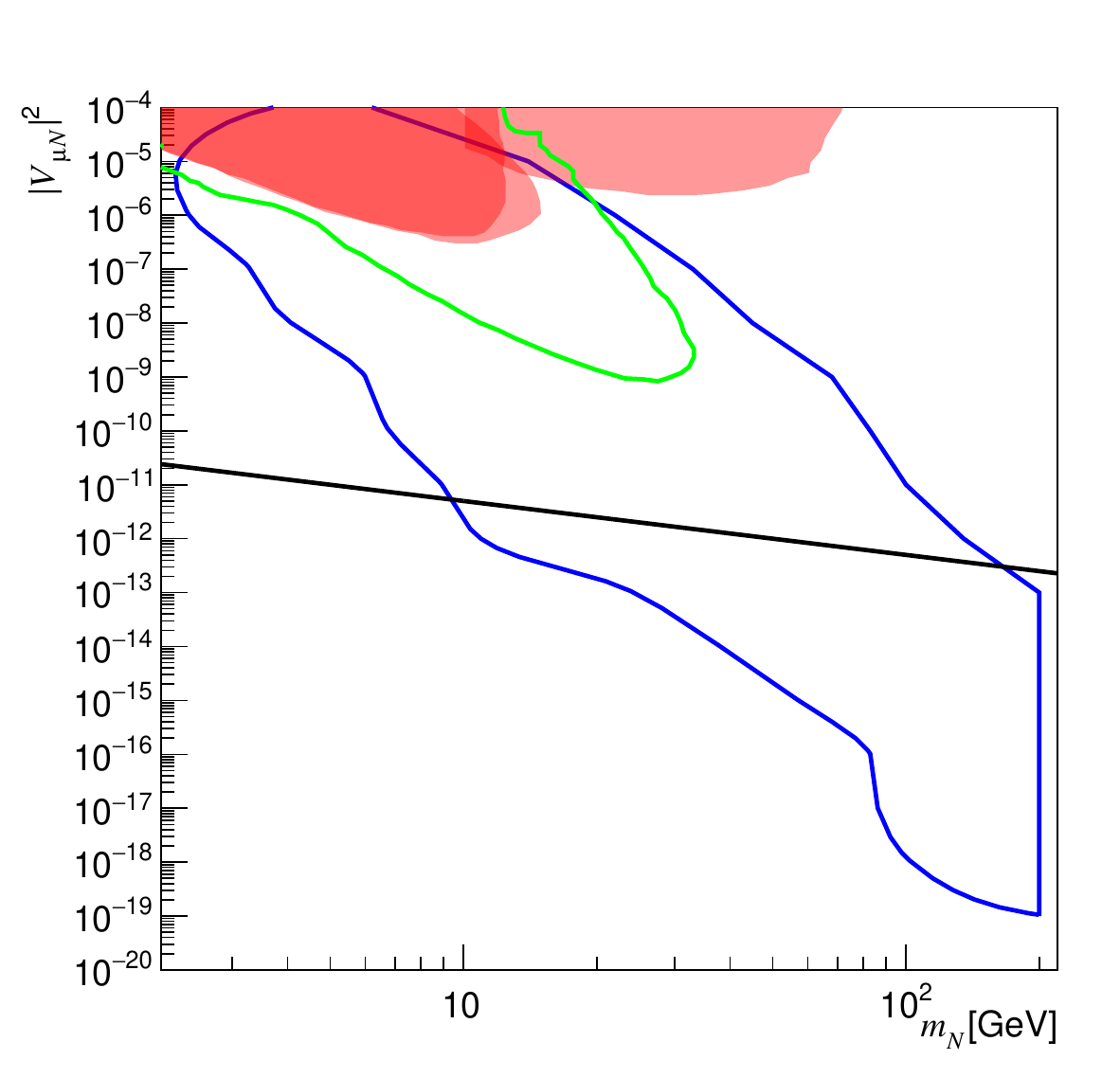}
		\includegraphics[width=0.45\linewidth]{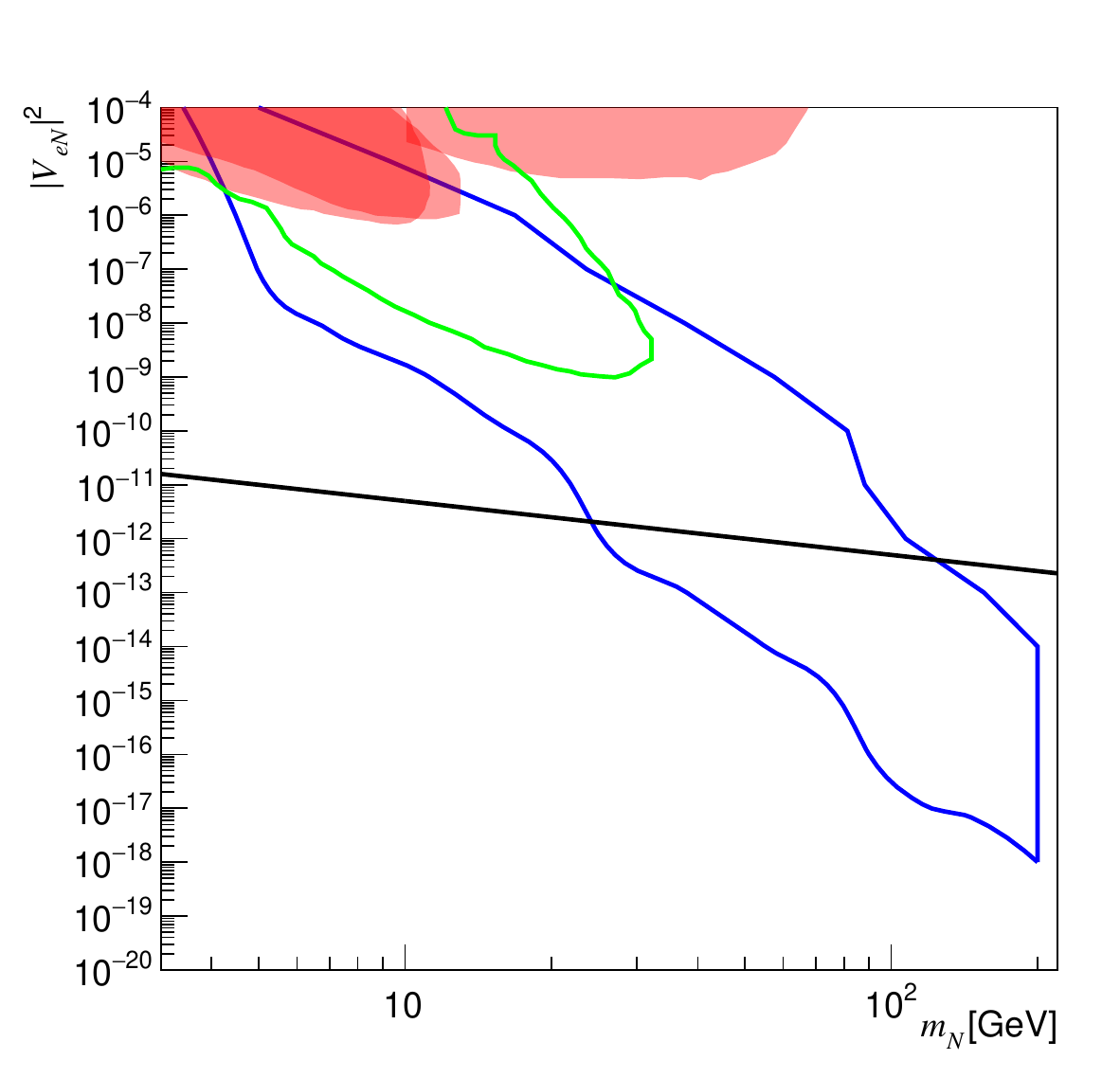}
		\includegraphics[width=0.45\linewidth]{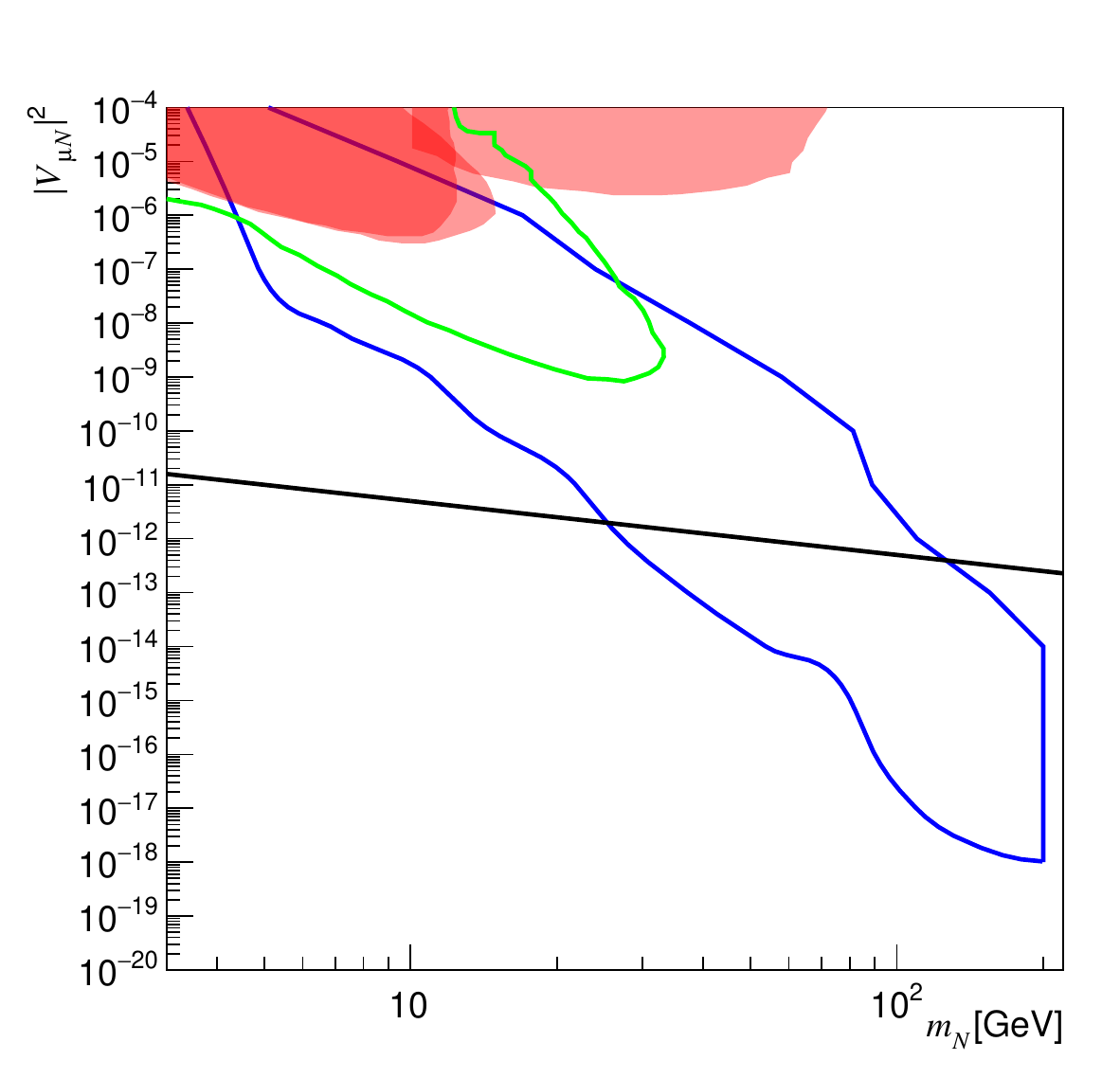}
	\end{center}
	\caption{Sensitivity reach of the 14 TeV LHC with luminosity $\mathcal{L} = 3ab^{-1}$ for a fixed value of $m_{H^+}=200$ GeV. The upper two panels are for one DV signature, and the lower two are for two DVs signature. The black lines are the natural seesaw value $V_{\ell N}^2\sim m_\nu/m_N$, where we have fixed $m_\nu=0.05$ eV for illustration. The red regions are excluded by direct search for HNL at LHC \cite{CMS:2023jqi,CMS:2024ake,CMS:2024xdq}. The green lines are the projected sensitivity reach of HL-LHC in the $W^\pm\to\ell^\pm N$ channel\cite{Drewes:2019fou}. In the left panels HNL $N$ only couples to electron and in the right panels HNL $N$ only couples to muon.}
	\label{scan01}
\end{figure}

Based on the above selection cuts, we first analyze the events with at least one DV. The results for a fixed value of $m_{H^\pm}=200$ GeV are shown in the upper two panels of Figure \ref{scan01}. It is clear that the DV signature can be discovered within the parameter space of 3 GeV $<m_N<200$ GeV when $10^{-19}<|V_{lN}|^2<10^{-4}$. Since the production of HNL is not suppressed by the mixing parameter $V_{\ell N}$, we can explore the range of $|V_{\ell N}|^2$ below seesaw predicted value. Compared with the canonical $W^\pm\to \ell^\pm N$ channel, the viable parameter space of the charged scalar $H^\pm\to\ell^\pm N$ channel is much larger. For instance, the $W^\pm\to \ell^\pm N$ channel could only probe $|V_{\ell N}|^2\gtrsim10^{-9}$ with $m_N\lesssim 30$ GeV at LHC \cite{Drewes:2019fou}, while the $H^\pm\to \ell^\pm N$ channel could probe $|V_{\ell N}|^2\gtrsim10^{-19}$ with $m_N< 200$ GeV. This upper limit corresponds to the kinetic threshold of $H^\pm\to \ell^\pm N$ decay, since we fix $m_{H^\pm}=200$ GeV in the analysis. The displacement cut $l_0\in[5,3000]$~mm leads to the upper and lower bound on $|V_{\ell N}|^2$ for certain $m_N$. Qualitatively speaking, the larger the $m_N$ is, the smaller the $|V_{\ell N}|^2$ would be. For $m_N=10$ GeV, the promising range is $10^{-11}\lesssim|V_{\ell N}|^2\lesssim10^{-5}$, which just reaches the seesaw predicted limit. For $m_N=100$ GeV, the promising region becomes $10^{-18}\lesssim|V_{\ell N}|^2\lesssim10^{-10}$. There is a dip around $m_N\sim m_W$, because the two-body decays as $N\to\ell^\pm W^\mp$ is allowed when $m_N>m_W$, so a much smaller $|V_{\ell N}|^2$ is required to satisfy the displacement cut of $l_0$. The sensitivity region of electron mixing and muon mixing patterns are quite similar. Although the acceptance rates of electron and muon at the detector are slightly different, it is shaded by the relatively large uncertainty of the Monte Carlo simulation procedure.

We then tighten the selection cuts by requiring at least two DVs in the final state while keeping the other cuts in Table \ref{Tab:cut} unchanged. The results are shown in the lower two panels of Figure~\ref{scan01}. Considering that the acceptance rate for DVs search at LHC detectors decreases linearly with length \cite{Alimena:2019zri,Knapen:2022afb}, we can simply assume that the reconstruction efficiency for two DVs signature search would decrease quadratically. Therefore, the overall discovery region of two DVs is $10^{-18}<|V_{\ell N}|^2<10^{-4}$ with HNL mass dependence, which is slightly smaller than that of one DV. In contrast to the one DV signature, the two DVs signal is free from the canonical $W^{\pm}\to\ell^\pm N$ background. For $m_N\sim$ few GeV with $|V_{\ell N}|^2\gtrsim10^{-5}$, the two DVs channel is better than the one DV channel to probe the new contribution from neutrinophilic scalars. However, such a region is now already excluded by LHC direct search.

\begin{figure}
	\begin{center}
		\includegraphics[width=0.45\linewidth]{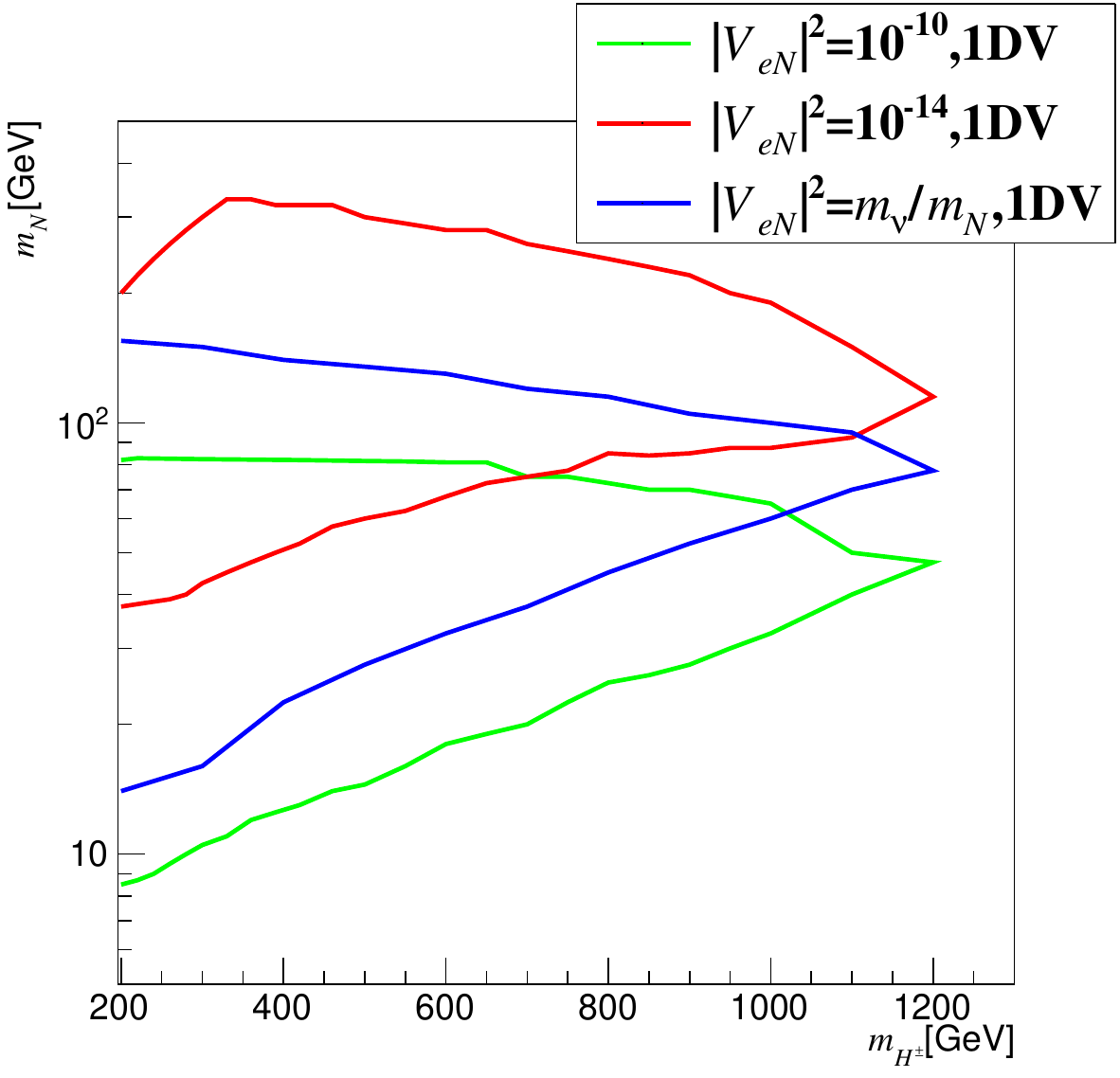}
		\includegraphics[width=0.45\linewidth]{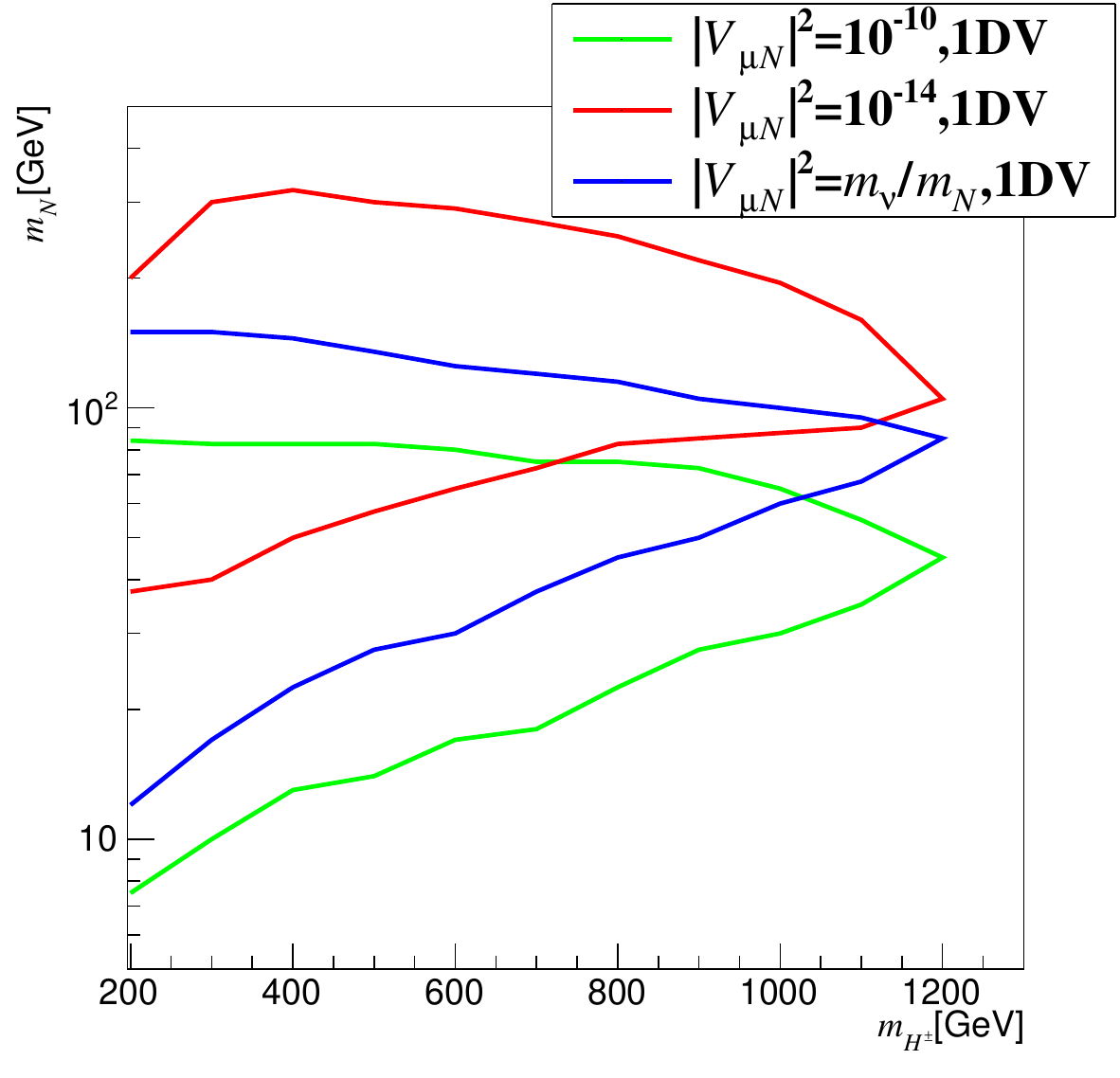}
		\includegraphics[width=0.45\linewidth]{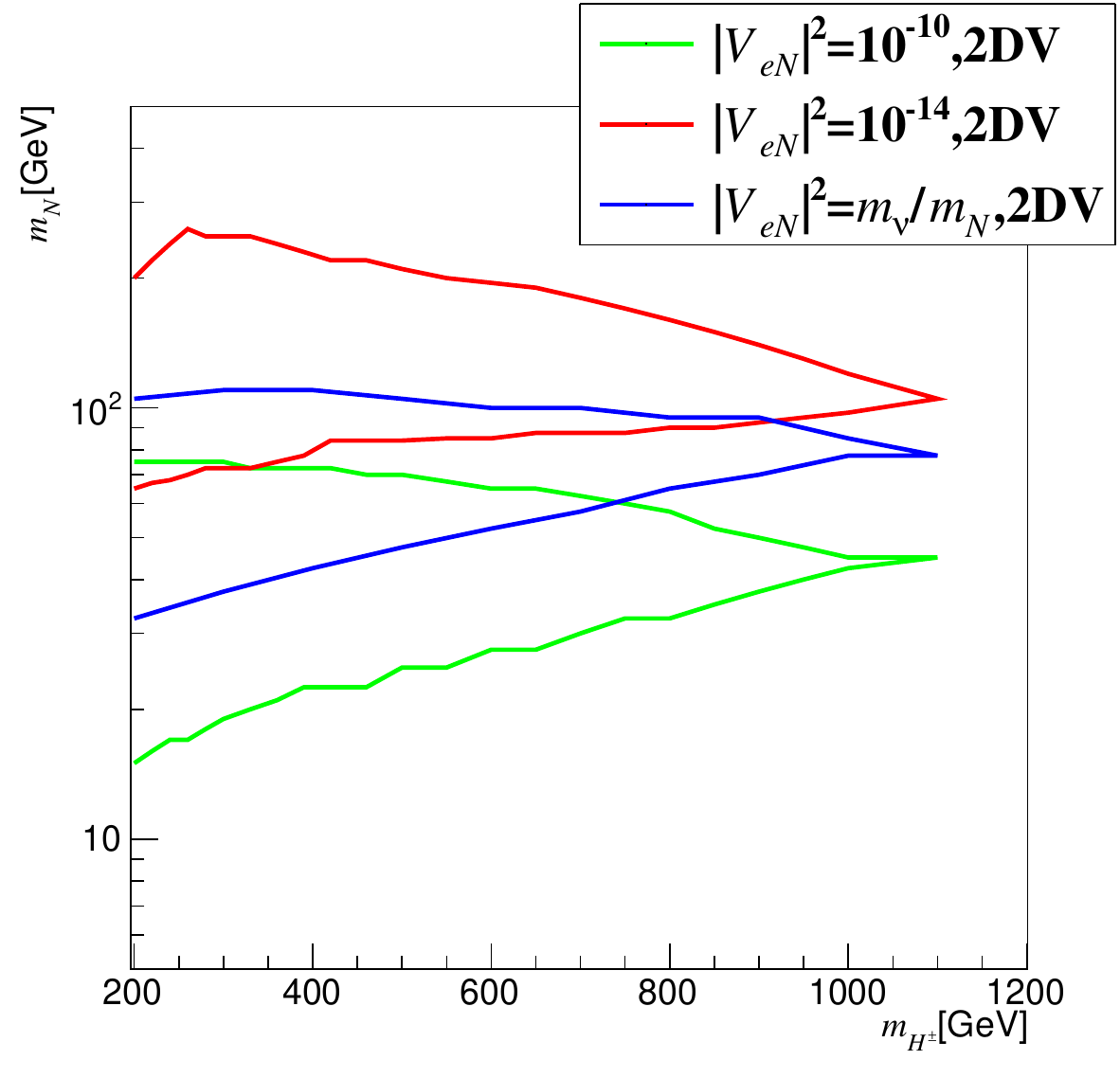}
		\includegraphics[width=0.45\linewidth]{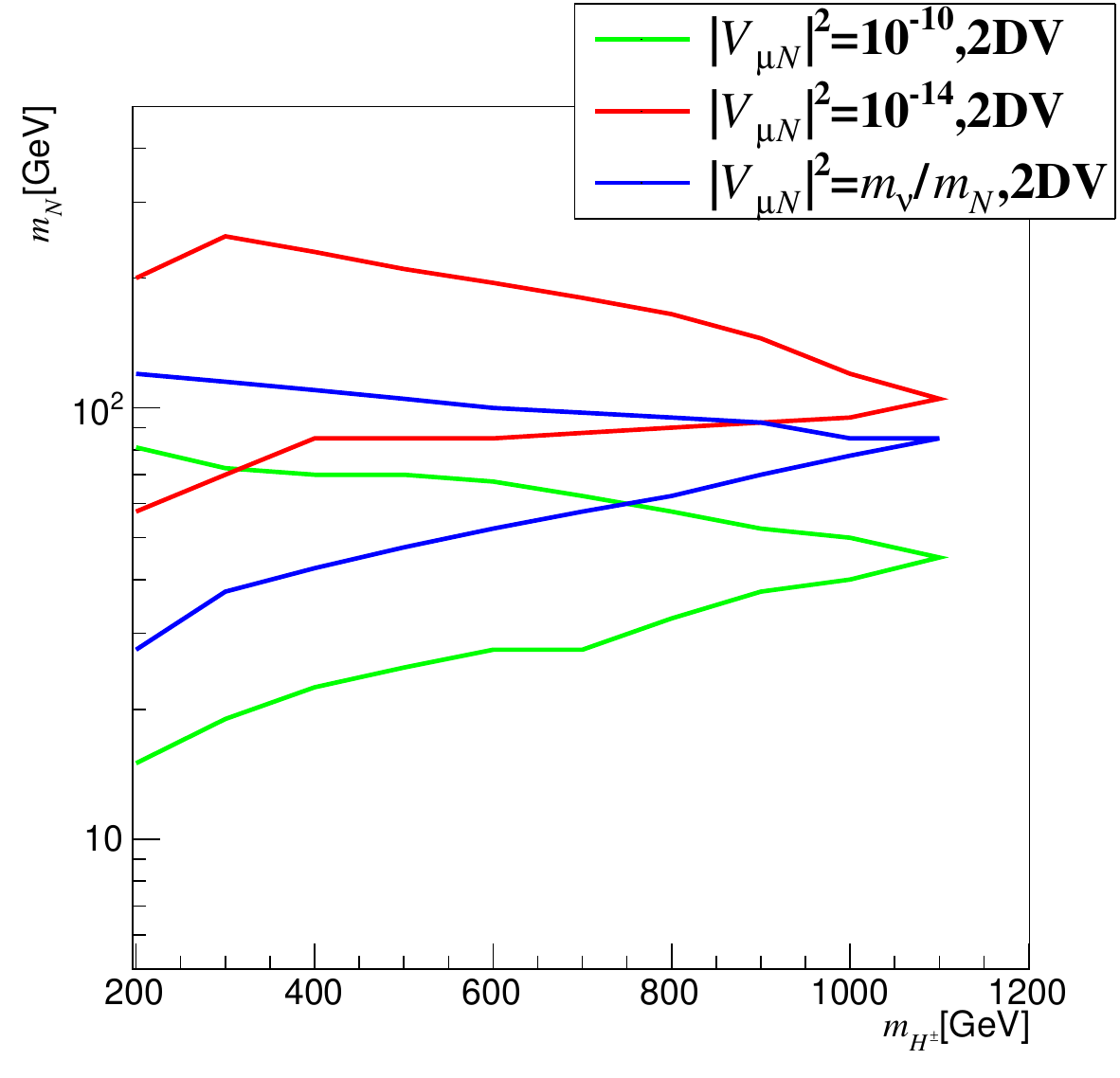}
	\end{center}
	\caption{Sensitivity reach of the 14 TeV LHC with luminosity $\mathcal{L} = 3~ab^{-1}$ for $|V_{\ell N}|^2=10^{-10},10^{-14}, m_\nu/m_N$. We have fixed $m_\nu=0.05$ eV for  calculation.}
	\label{scan03}
\end{figure}

Besides the HNL mass $m_N$ and mixing parameter $V_{\ell N}$, the cross section of the DV signature also depends on the charged scalar mass $m_{H^+}$. In Figure \ref{scan01}, it is shown that the upper bound of $m_N$ for the DV signature corresponds to the charged scalar mass $m_{H^+}$ with the proper mixing parameter $V_{\ell N}$. By setting the mixing parameter $|V_{lN}|=10^{-10}$ and $|V_{lN}|=10^{-14}$, we then explore the promising region of the DV signature in the $m_N-m_{H^+}$ plane at the HL-LHC. The benchmark value $|V_{\ell N}|=10^{-10}$ is larger than the canonical seesaw prediction value $|V_{\ell N}|^2\simeq m_\nu/m_N\sim10^{-12}$, meanwhile the benchmark value $|V_{\ell N}|=10^{-14}$ is smaller than the theoretical favor prediction. The results are shown in Figure~\ref{scan03}. For one DV signature, the sensitive region of $m_{H^+}$ can reach 1200 GeV. Due to lower reconstruction efficiency, the sensitive region of $m_{H^+}$ for two DV signal would reach about 1100 GeV. A larger value of $m_{H^+}$ will lead to a smaller sensitive region of $m_N$. For example, the sensitive region is $m_N\in[10,80]$ GeV when $m_{H^+}=300$ GeV and $|V_{\ell N}|^2=10^{-10}$, which reduces to about $m_N\in[20,70]$ GeV when $m_{H^+}=700$~GeV.  Considering similar tagging efficiency for electron and muon at LHC, the pure muon mixing won't give essentially different results from the electron mixing pattern.

For the promising region of $m_N$, it heavily depends on the mixing pattern. A smaller mixing parameter $V_{lN}$ usually requires a larger $m_{N}$ to satisfy the displacement cut $5~{\rm{mm}}<l_0<3000~\rm{mm}$. When $|V_{\ell N}|^2=10^{-10}$, the promising regions always satisfy $m_N<m_W$ for both one DV and two DVs signature, which indicates that the three body decays of HNL $N$ are the dominant decay modes for mixing parameter $|V_{\ell N}|^2$ larger than the seesaw predicted value. On the other hand, when $|V_{\ell N}|^2(=10^{-14})$ is smaller than the theoretical favor value, most of promising regions of DV signatures would satisfy $m_N>m_W$, so two body decays of HNL $N$ become the dominant channels. In this scenario with $|V_{\ell N}|^2=10^{-14}$, one may probe $m_N$ up to $\sim300$ GeV. 

The most natural scenario is the seesaw predicted value $|V_{\ell N}|^2=m_\nu/m_N$. The results are also shown in Figure \ref{scan03}. As already shown in Figure \ref{scan01}, current DV searches via the $W^\pm\to \ell^\pm N$ channel can not probe such tiny mixing parameter. With an unsuppressed cross section, we find that the $H^\pm\to\ell^\pm N$ channel could probe $m_{H^+}\lesssim1200$ GeV in this scenario. The promising range of $m_N$ heavily depends on $m_{H^{+}}$. Typically for $m_{H^+}=600$ GeV, we could probe $m_N\in[30,110]$ GeV in the one DV channel and $m_N\in[50,100]$ GeV in the two DVs channel.

\subsection{Signature at CLIC }

\begin{figure}
	\begin{center}
		\includegraphics[width=0.45\linewidth]{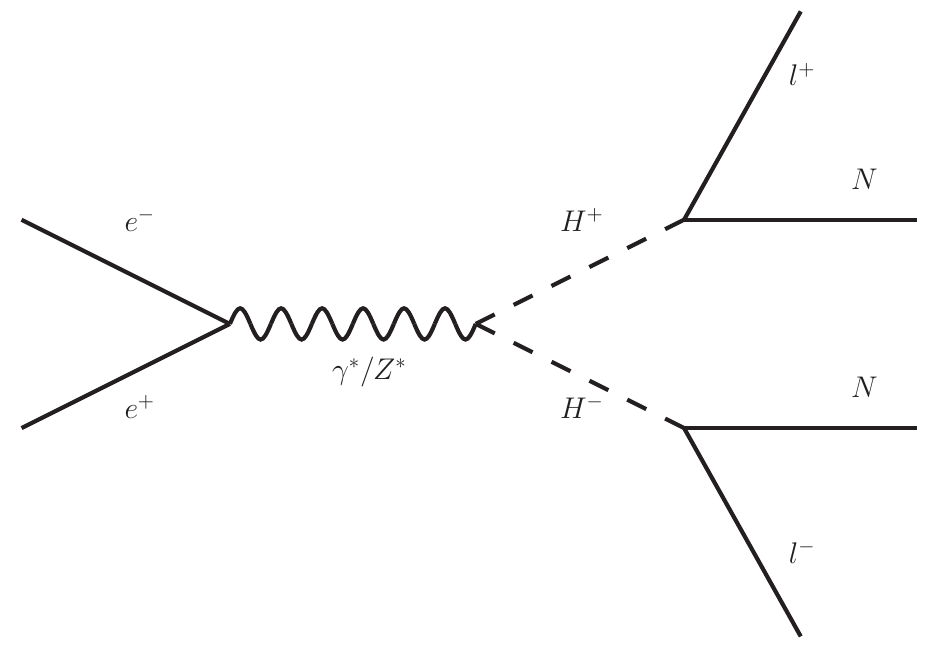}
		\includegraphics[width=0.36\linewidth]{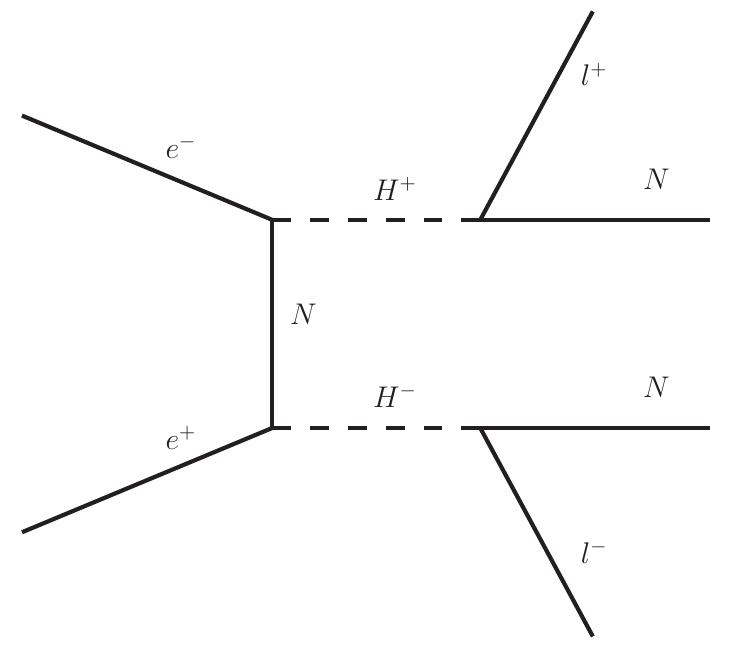}
	\end{center}
	\caption{Feynman diagrams for the ${e^+}{e^-}\rightarrow{H^+}{H^-}\rightarrow{\ell^+}N \ell^-N $ processes.}
	\label{ee}
\end{figure}

The Compact Linear Collide (CLIC)\cite{ILC:2007oiw,CLICdp:2018cto,Brunner:2022usy} is a proposed multi-TeV $e^+e^-$ linear collider. In this paper, we consider the 3-TeV collision energy stage with a high luminosity of $\mathcal{L} = 5~{\rm ab}^{-1}$.  According to the results in Figure \ref{cs}, the cross section of $H^+H^-$ at CLIC is much larger than it is at LHC for TeV scale $m_{H^+}$. So CLIC is expected more promising to probe the heavily charged scalar region. In the $\nu$2HDM model, the long-lived HNL can be generated at CLIC through ${e^+}{e^-}\rightarrow{H^+}{H^-}\rightarrow{\ell^+}N \ell^-N $ process. As shown in Figure \ref{ee}, there are two different production channels for $H^+H^-$. The $s$-channel process is mediated by virtual $\gamma^*$ or $Z^*$, which only depends on the mass of the charged scalar. On the other hand, the $t$-channel process is mediated by the HNL $N$, which is also determined by the Yukawa coupling $y$.  In principle, large $y_{eN}$ can be obtained by tunning the structure of Yukawa matrix $y$. For simplicity, we consider $y\sim10^{-2}$ to avoid tight constraints from the lepton flavor violation. In this way, the contribution of the $t$-channel process can be neglected. So the following results of CLIC in this study are conservative.

Similar to the LHC study, we also focus on the decay of charged scalar $H^\pm\to\ell^\pm N$ to distinguish the $\nu$2HDM from other HNL models in this section. Therefore, the DVs signature from $e^+e^-\to HA\to \nu_\ell N \nu_\ell N$ is not taken into account. It should be mentioned that the DVs signature also arises from process $e^+e^-\to NN$ via $t$-channel mediator of $H^\pm$. However, the cross section of $e^+e^-\to NN$ is also suppressed by small Yukawa coupling $y\sim10^{-2}$ in our consideration. On CLIC, searches for long-lived particles through DVs signature have no irreducible SM background.  There is one irreducible  background from new physics as $e^+e^-\to W^+W^-\to \ell^+ N\ell^- N$. However, the corresponding cross section is suppressed by the mixing parameter as $|V_{\ell N}|^4$, which is negligible under current experimental limits.  So it is still safe to consider the total backgrounds as one.

\begin{figure}
	\begin{center}
		\includegraphics[width=0.45\linewidth]{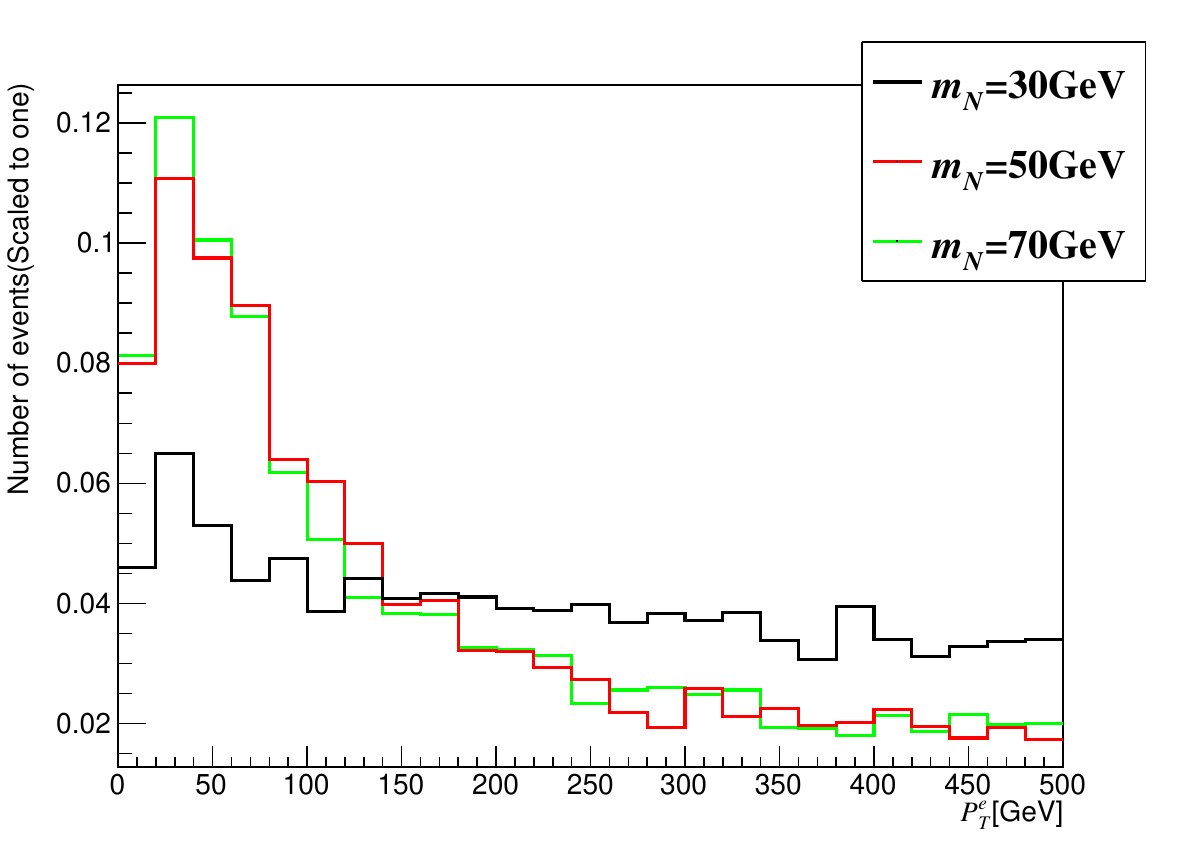}
		\includegraphics[width=0.45\linewidth]{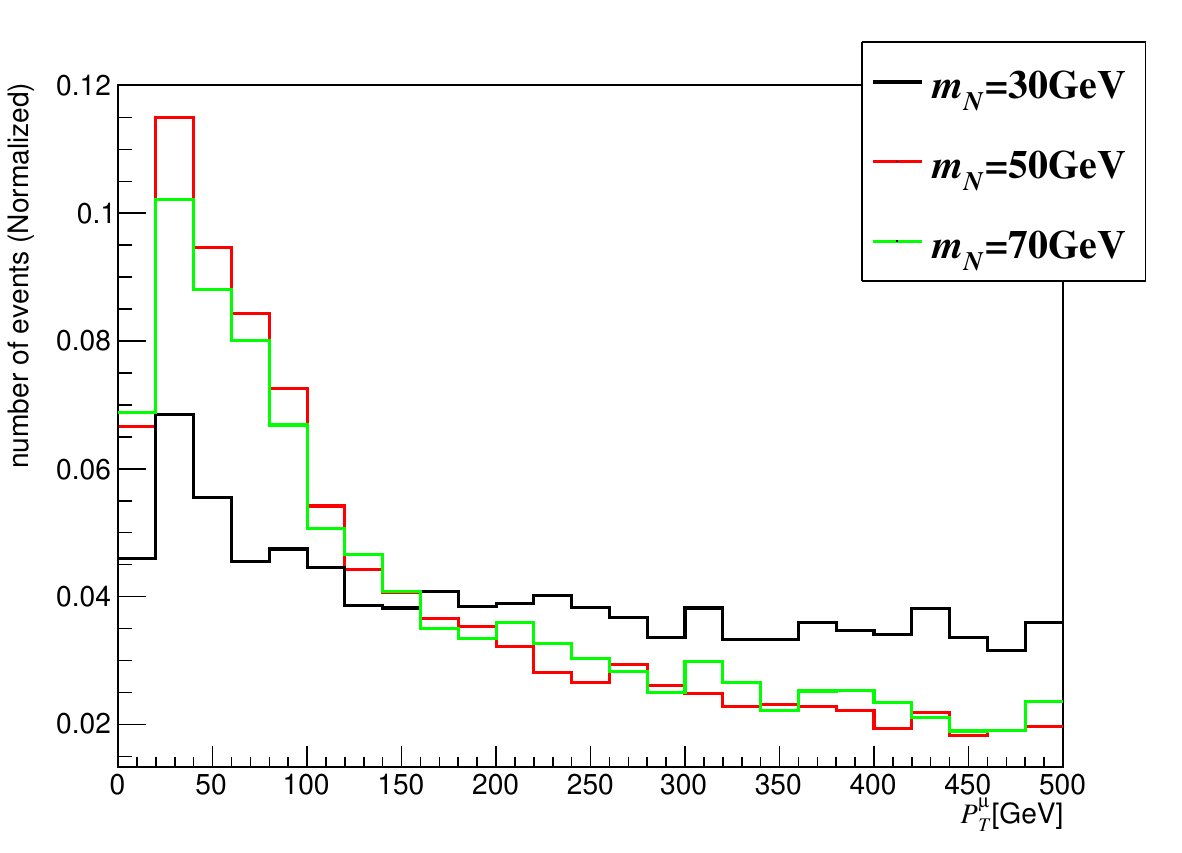}
		\includegraphics[width=0.45\linewidth]{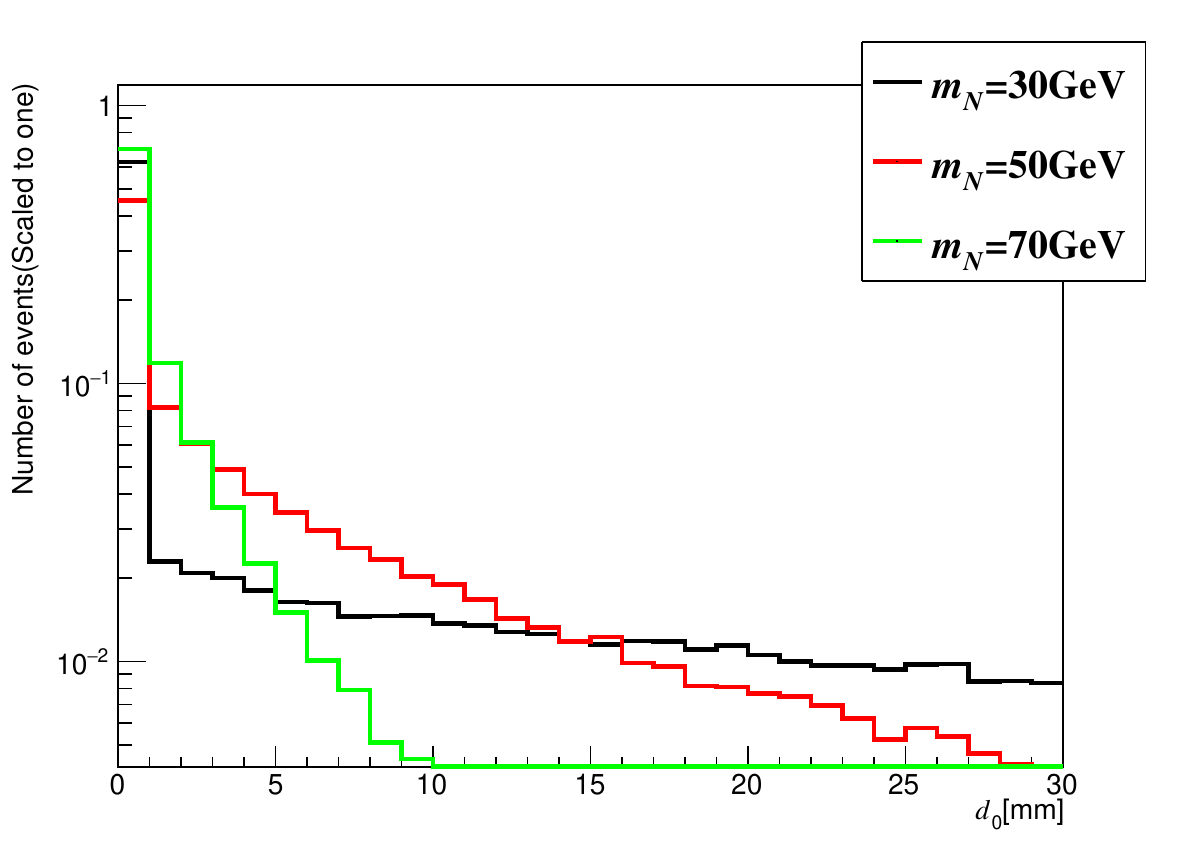}
		\includegraphics[width=0.45\linewidth]{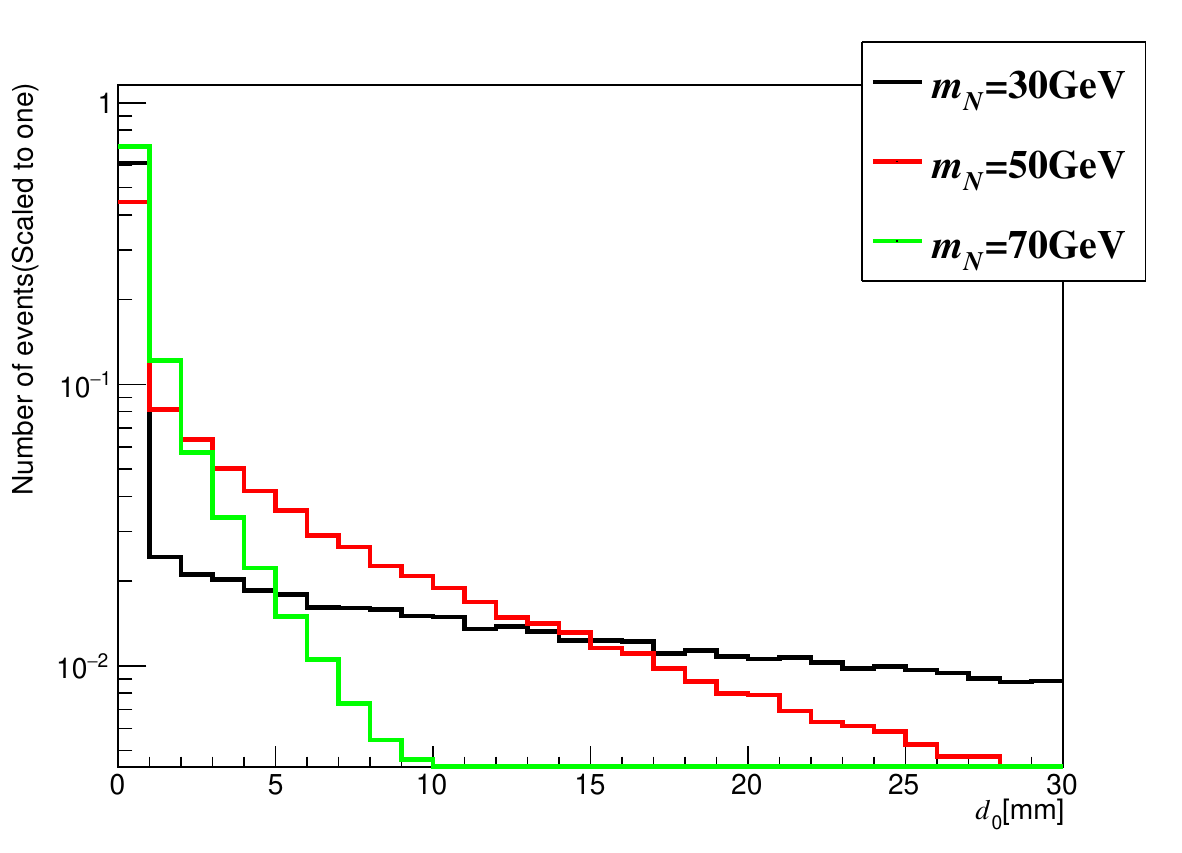}
		\includegraphics[width=0.45\linewidth]{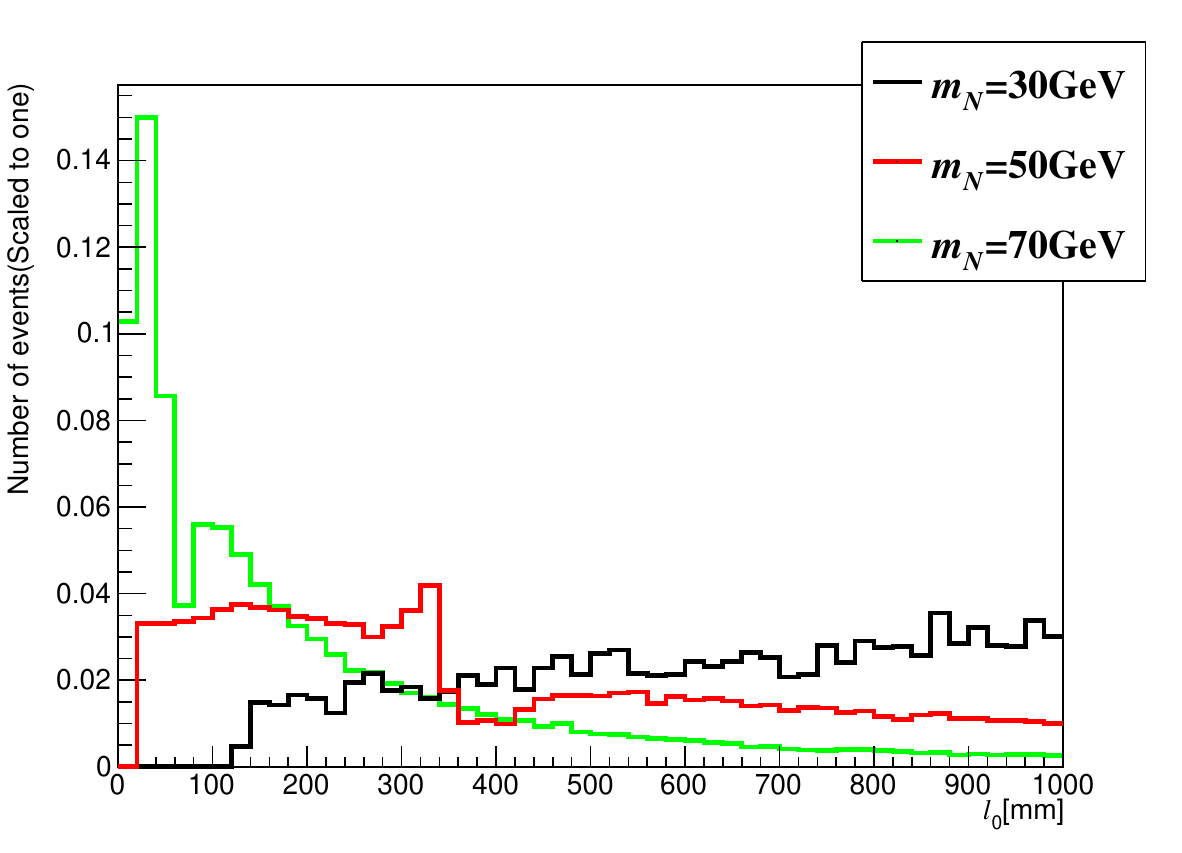}
		\includegraphics[width=0.45\linewidth]{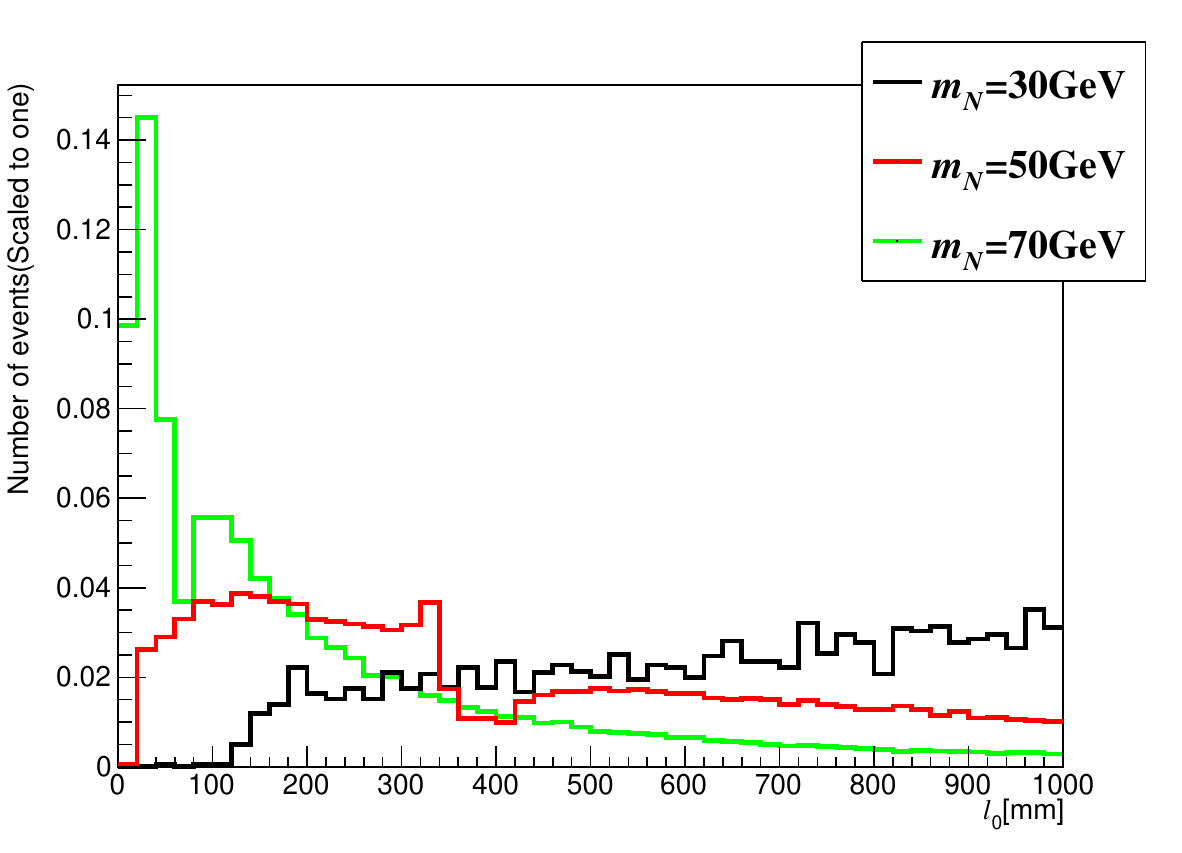}
		\includegraphics[width=0.45\linewidth]{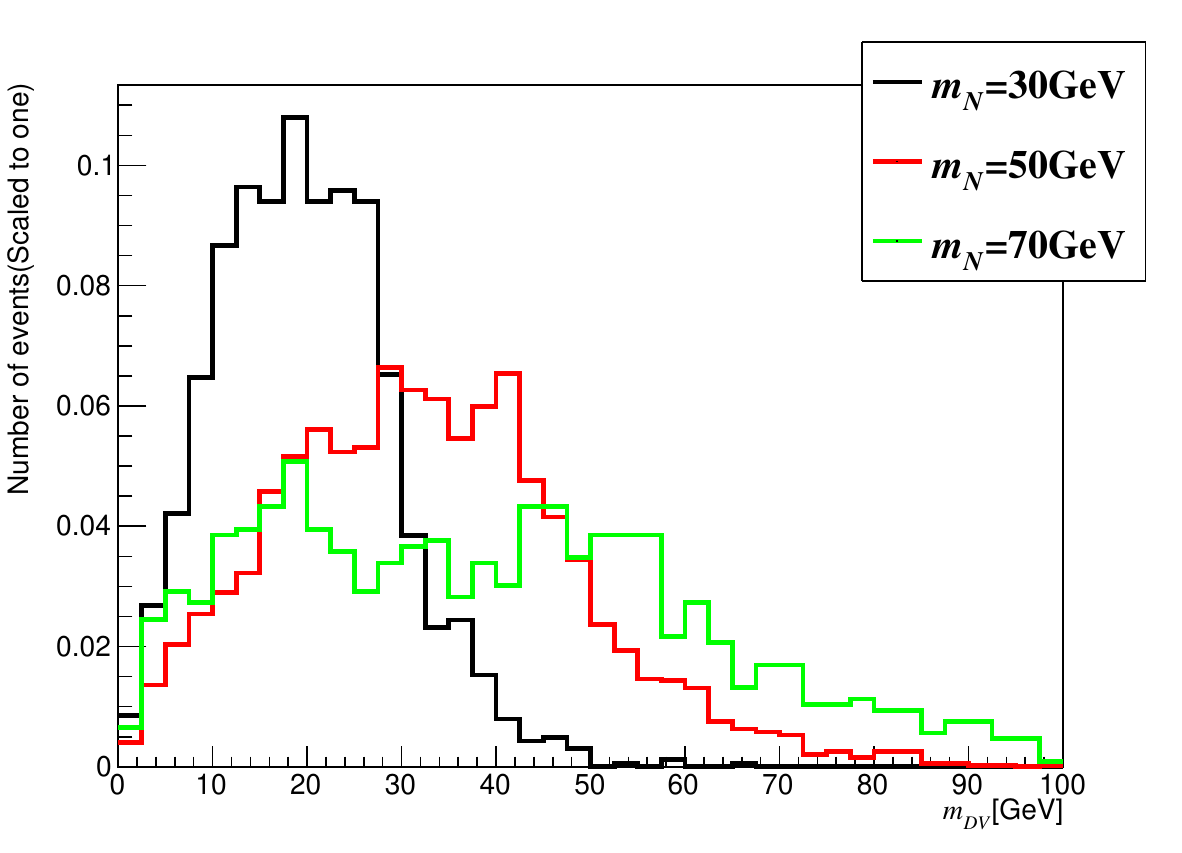}
		\includegraphics[width=0.45\linewidth]{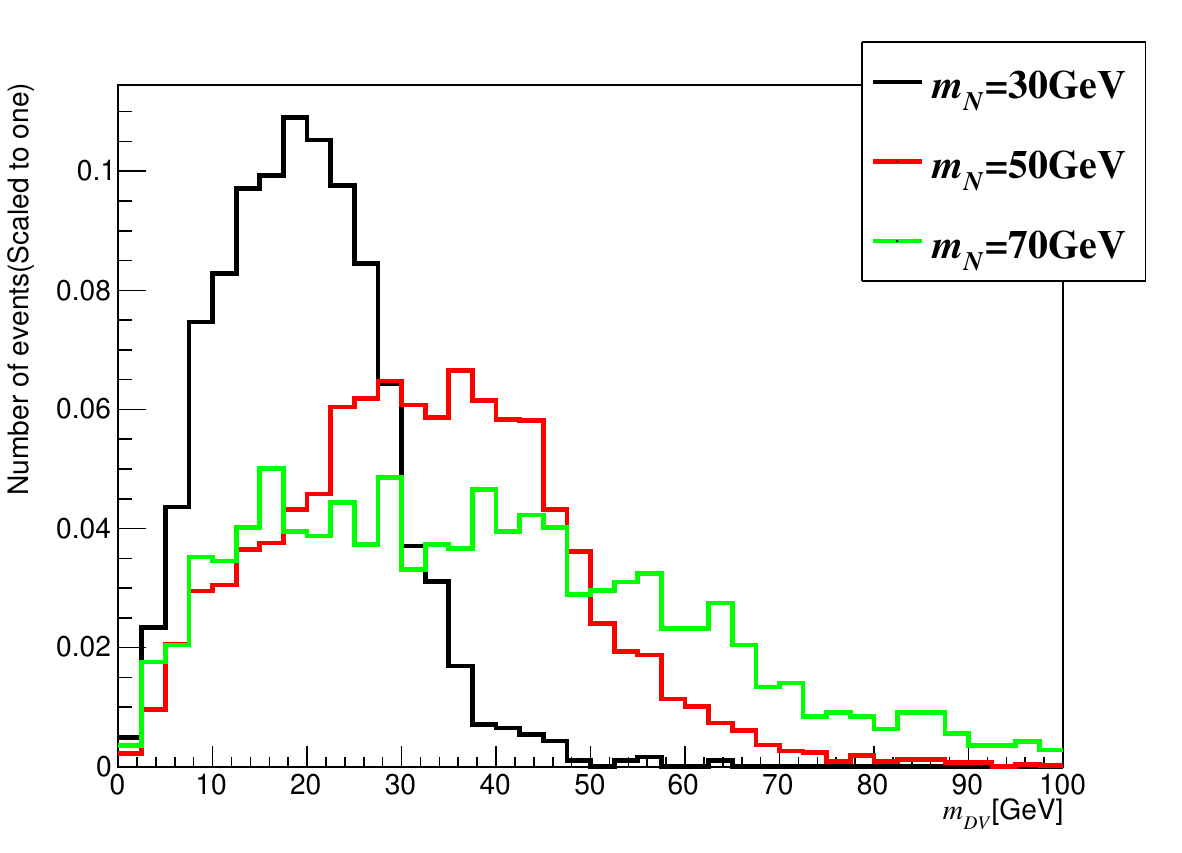}	
	\end{center}
	\caption{Same as Fig.\ref{track}, but for 3 TeV CLIC.}
	\label{clic_track}
\end{figure}

To search for the possible existence displaced vertex signature on CLIC, we perform a similar analysis as on LHC. In Figure~\ref{clic_track}, we show distributions to use. As before, three benchmark points are selected as $m_{N}=30,50,70$GeV with $m_{H^+}=200$ GeV and $|V_{\ell N}|^2=10^{-10}(\ell=e ~\text{or}~\mu)$.  Considering the clean environment for DVs search at CLIC, we did not put any backgrounds as a reference. Distributions of electron mixing and muon mixing are quite similar to each other. Benchmark points can be distinguished by variables as $P_T^\ell,d_0,l_0,m_{DV}$. As different detector geometries of CLIC  and LHC, we have changed the selection cuts on displacement as $5~{\rm{mm}}<l_0<1100~\rm{mm}$, meanwhile other cuts are kept unchanged.  We summarize all of the selection cuts for CLIC in Table~\ref{Tab:clic_cut}.

\begin{table}
	\begin{center}
		\begin{tabular}{|c | c | c | c | c|} 
			\hline
			\hline
			& \multicolumn{4}{c|}{\textbf{Cut-Flow}}  \\
			\hline
			\textbf{Trigger on Lepton} & \multicolumn{4}{|c|}{$N_\ell\geq1,$ $P_T^\ell>20\rm{GeV},$ $|\eta_\ell|<2.5$}  \\
			\hline
			\textbf{Tracks}  & \multicolumn{4}{|c|}{$P^{trk}>5\rm{GeV},$ $d_0>2\rm{mm}$} \\
			\hline
			\textbf{Displaced Vertex} & \multicolumn{4}{|c|}{$N_{\ell}\geq1,$ $N_{trk}\geq2,$  $\Delta{x}<1{\rm{mm}},\Delta{y}<1{\rm{mm}},\Delta{z}<1{\rm{mm}}$}  \\
			\cline{2-5}
			\textbf{Reconstruction}& \multicolumn{4}{|c|}{$\Delta{R}>0.1,$ $l_0\in[5,1100]\rm{mm},$ $m_{DV}\geq5\rm{GeV}$}  \\
			\hline
			\hline
			Benchmark Point & 1 Displaced Vertex & Significance & 2 Displaced Vertex & Significance \\
			\hline
			$m_N=30$ GeV & 1.10(1.11) & 290(292) & 0.12(0.12)  & 82(82) \\
			\hline
			$m_N=50$ GeV & 2.01(2.02) & 407(408) & 0.53(0.52)  & 193(2) \\
			\hline
			$m_N=70$ GeV & 0.68(0.68) & 211(211)  & 0.04(0.04)  & 44(44) \\
			\hline
		\end{tabular}
	\end{center}
	\caption{Same as Table \ref{Tab:cut} but for the 3 TeV CLIC with luminosity $\mathcal{L} = 5~{\rm ab}^{-1}$. \label{Tab:clic_cut}}
\end{table}

For the three benchmark points of $m_N=$30 GeV, 50 GeV, 70 GeV, the significance could reach 290(292), 407(408), 211(211) for one DV pure electron(muon) mixing pattern, and 82(82), 193(190), 44(44) for two DVs pure electron(muon) mixing pattern. Qualitatively speaking, the electron channels have similar significance as the muon channels at CLIC. With $|V_{\ell N}|^2=10^{-10}$, the most promising benchmark point is $m_N=50$ GeV. Compared with LHC, the significance for benchmark points at CLIC are smaller, because of the smaller production cross section of $H^+H^-$ for $m_{H^+}=200$ GeV.

\begin{figure}
	\begin{center}
		\includegraphics[width=0.45\linewidth]{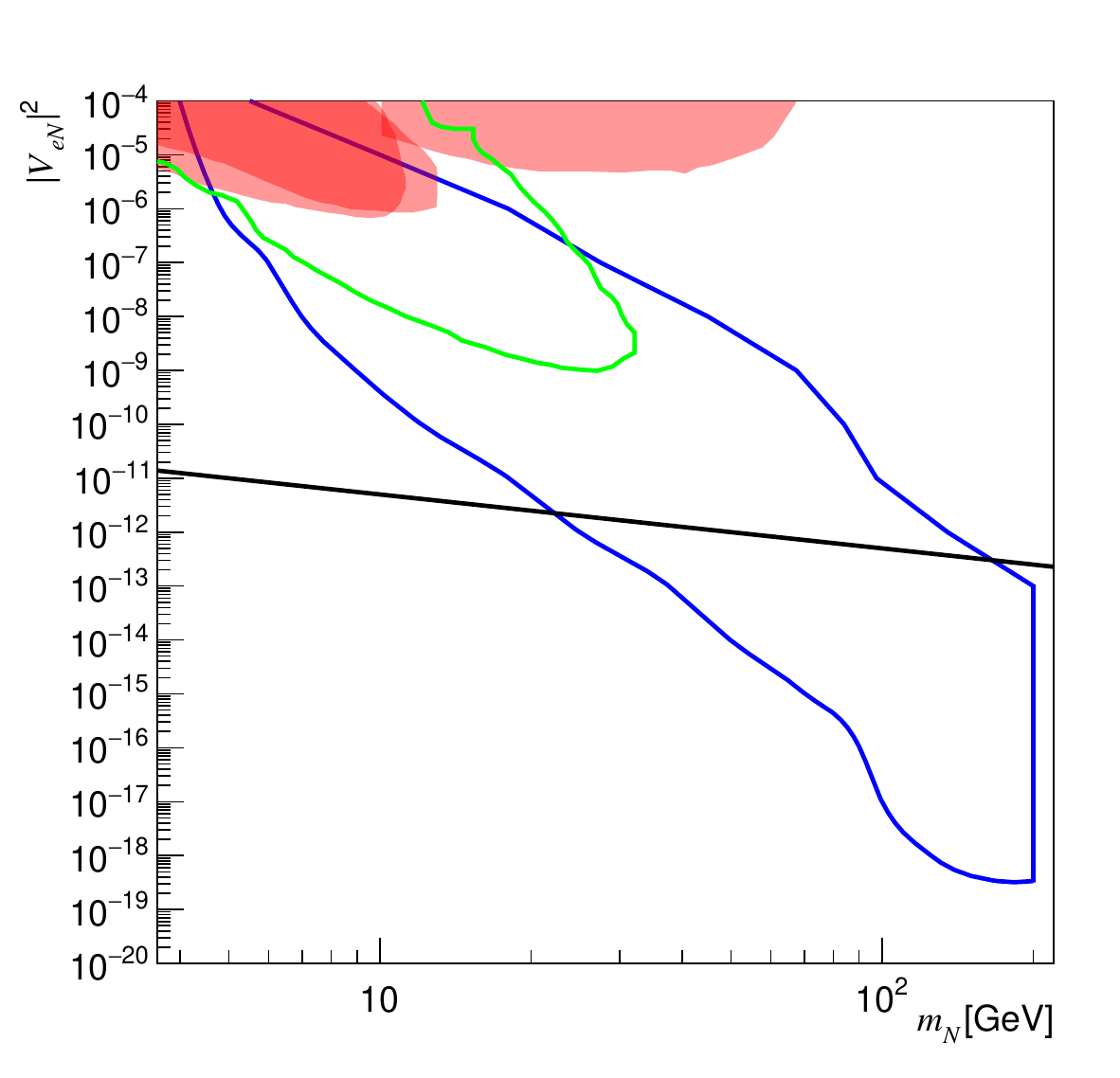}
		\includegraphics[width=0.45\linewidth]{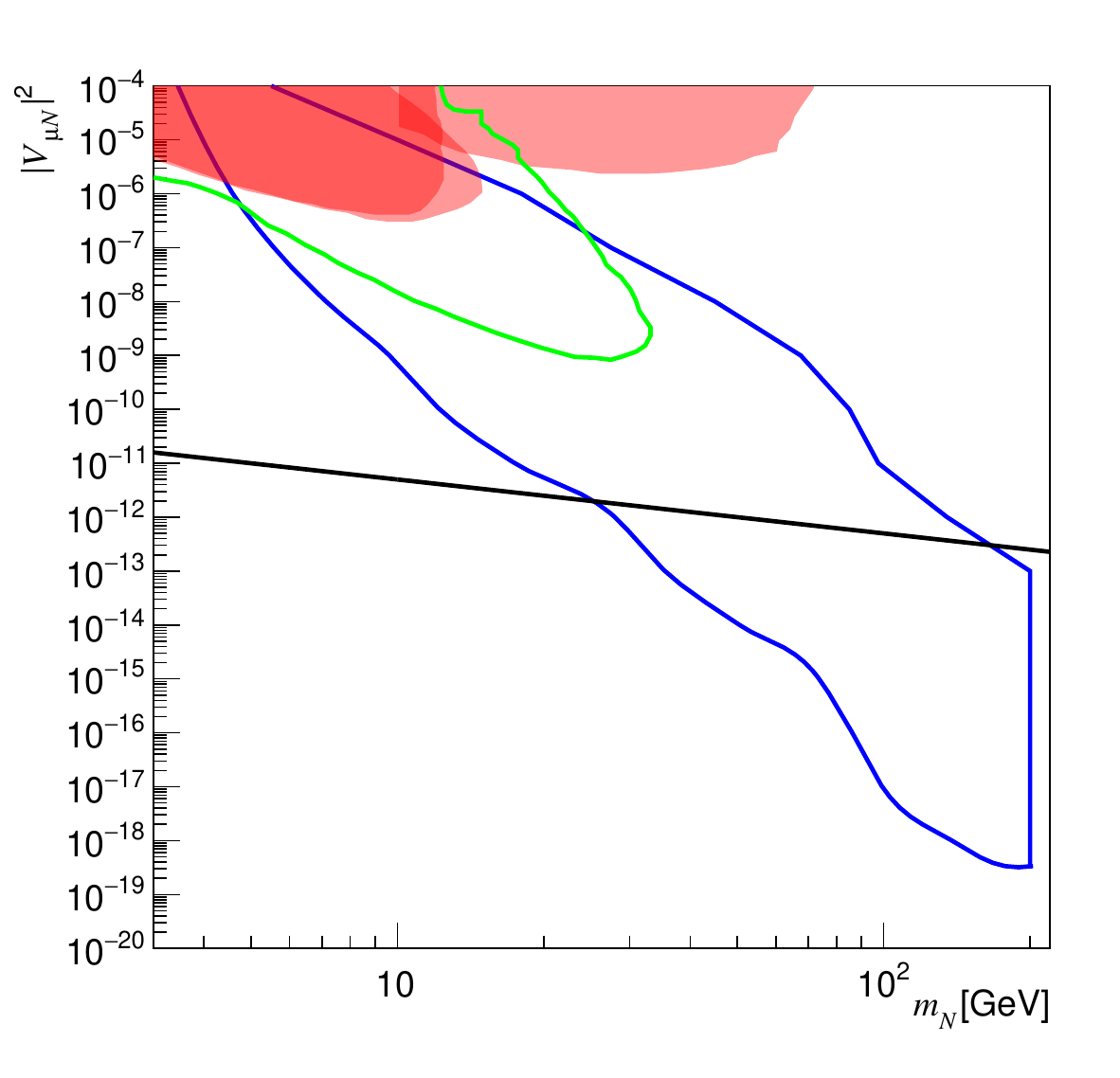}
		\includegraphics[width=0.45\linewidth]{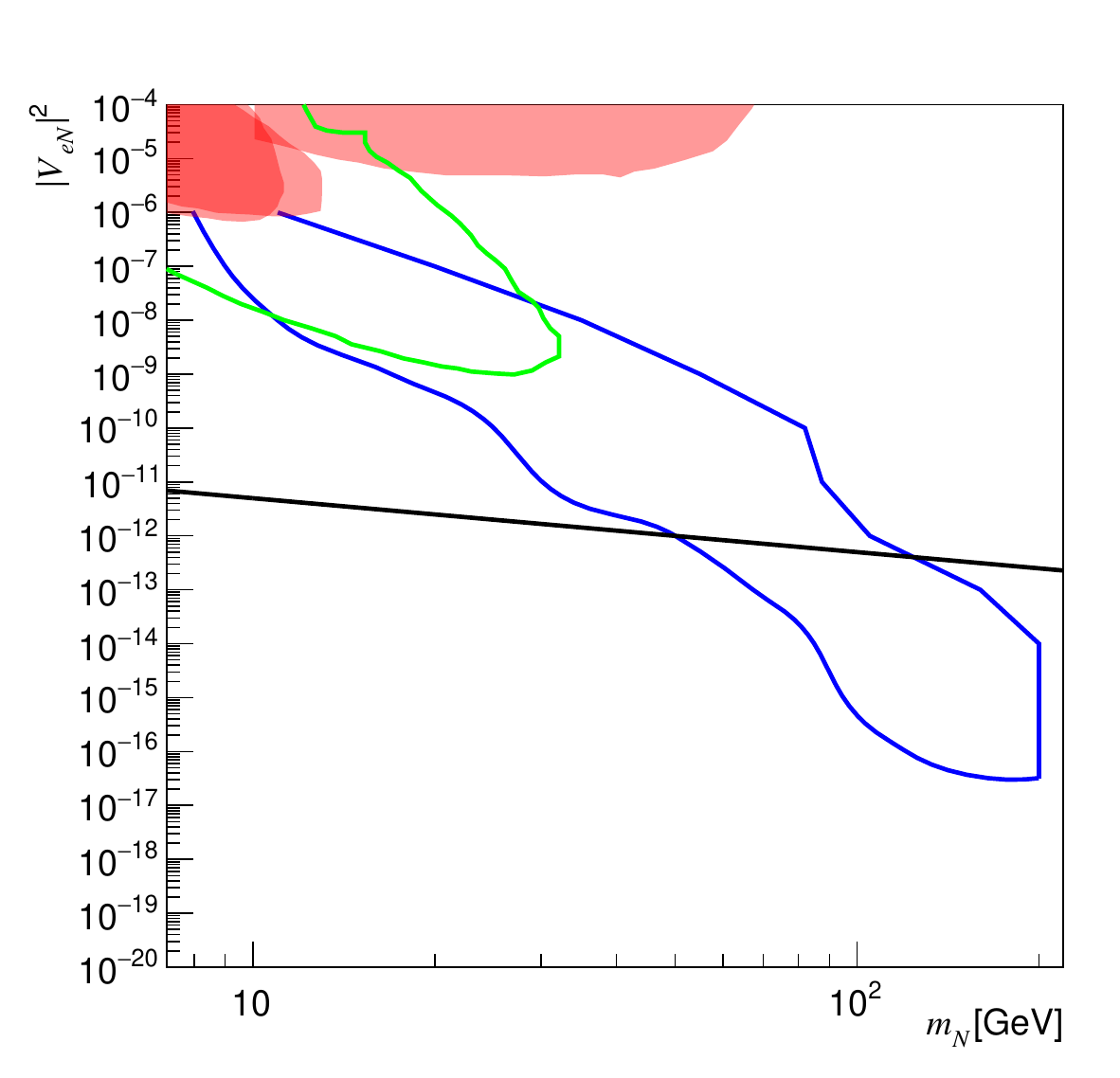}
		\includegraphics[width=0.45\linewidth]{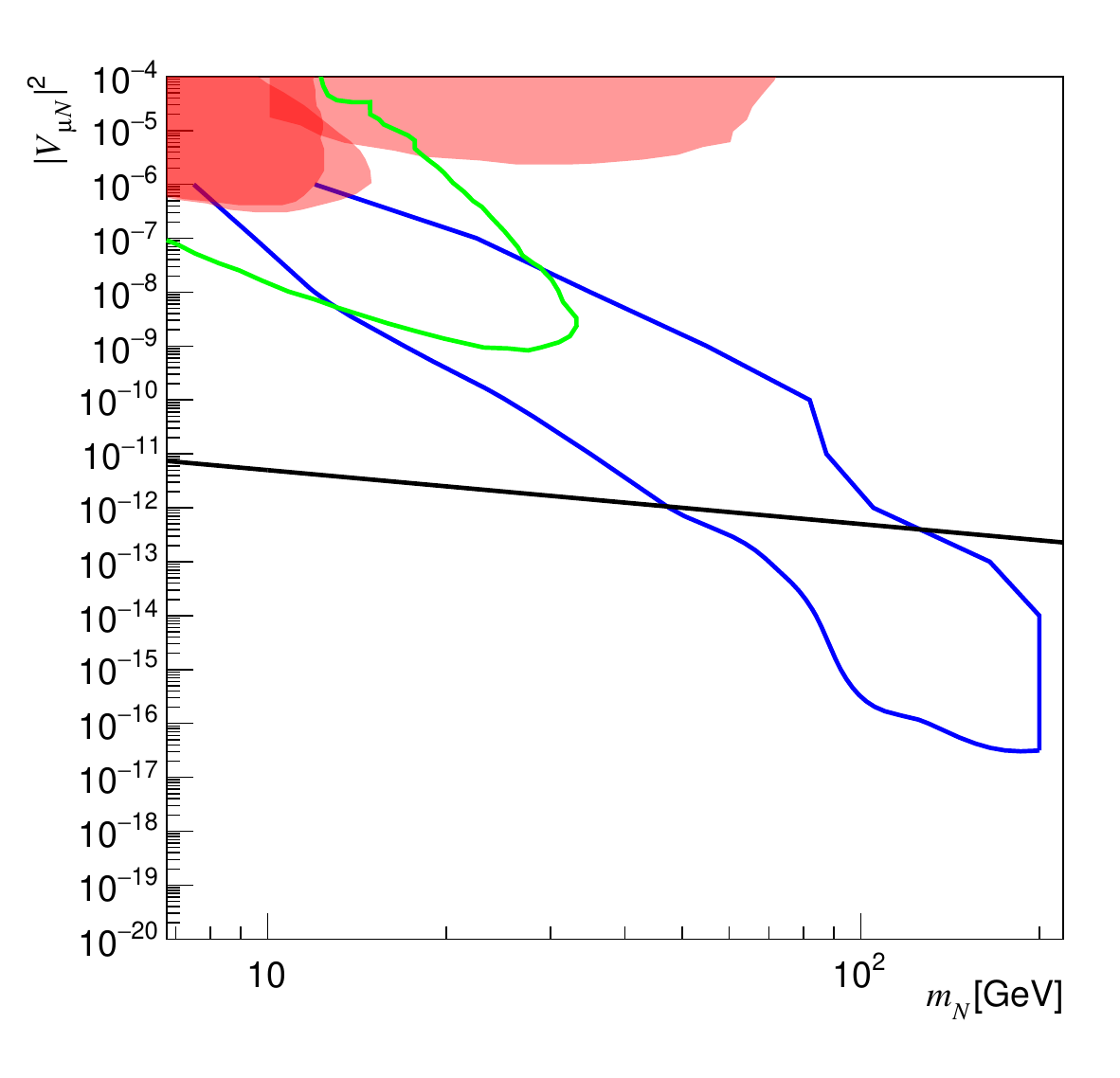}
	\end{center}
	\caption{Same as Fig.~\ref{scan01} but for sensitivity reach of the 3 TeV CLIC with luminosity $\mathcal{L} = 5~{\rm ab}^{-1}$. }
	\label{scan04}
\end{figure}

We then explore the parameter space for one DV signature with fixed charged scalar mass $m_{H^\pm}=200$~GeV. The results are shown in the upper two panels of Figure \ref{scan04}. In both the electron and the muon mixing pattern, we can probe $|V_{\ell N}|^2\gtrsim10^{-18}$ when $m_N>3.5$ GeV.  For seesaw induced mixing $|V_{\ell N}|^2=m_\nu/m_N$, the one DV signature could discover 20 GeV $\lesssim m_N\lesssim130$ GeV. Compared with the sensitive regions of LHC, the regions of CLIC are slightly smaller, because the cross section of $H^+H^-$ at CLIC is smaller than it is at LHC for $m_{H^\pm}=200$~GeV. Therefore, if no DV signature is discovered at LHC, then CLIC can hardly have any positive DV signature for light charged scalar.

For the two DVs signature, scanned results are shown in the lower two panels of Figure \ref{scan04}. Compared with the one DV signal, the promising areas become smaller due to more strict selection cuts. For pure electron or muon mixing pattern, we can probe HNLs with the square of the mixing parameter as small as $|V_{\ell N}|^2\sim10^{-16}$, and the mass as small as $m_{N}\sim8$ GeV.  In the narrow mass region between 40 GeV and 110 GeV, we may discover the two DV signature with seesaw induced mixing $|V_{\ell N}|^2=m_\nu/m_N$. We also find that the two DV signature at CLIC could not probe regions with $|V_{\ell N}|^2>10^{-6}$, which is different from LHC. According to Figure \ref{scan01}, $|V_{\ell N}|^2>10^{-6}$ requires $m_N\lesssim10$ GeV for observable DV signature. Due to the selection cut on the reconstructed DVs mass of $m_{DV}>5$ GeV, the acceptance efficiency for the DV signal is relatively low in this region. Meanwhile, the cross section of the signal process on 3 TeV CLIC is much smaller than on 14 TeV LHC for $m_{H^+}=200$ GeV, so the two DVs signature is not sensitive for mixing parameter $|V_{\ell N}|^2>10^{-6}$.

\begin{figure}
	\begin{center}
		\includegraphics[width=0.45\linewidth]{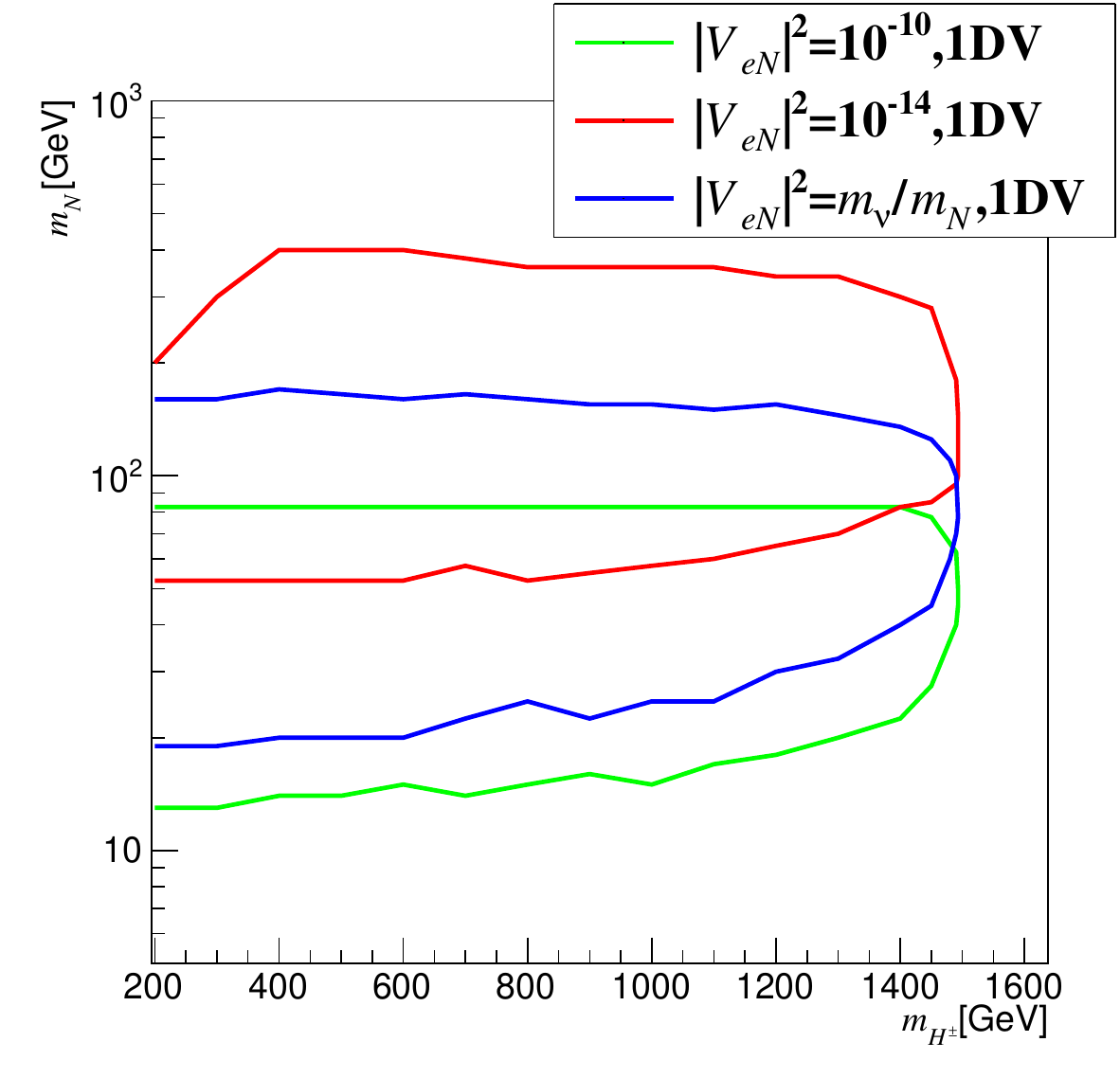}
		\includegraphics[width=0.45\linewidth]{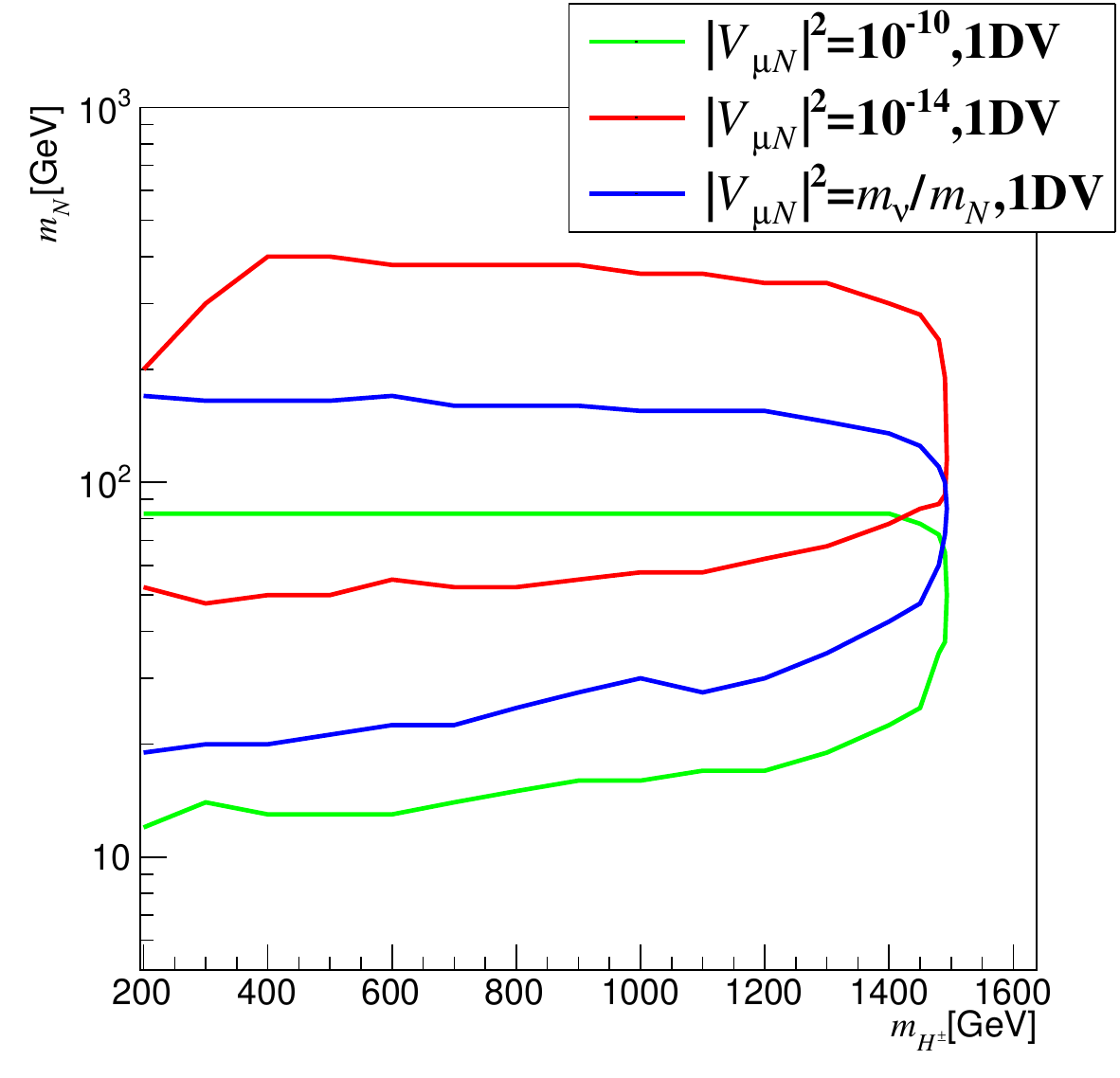}
		\includegraphics[width=0.45\linewidth]{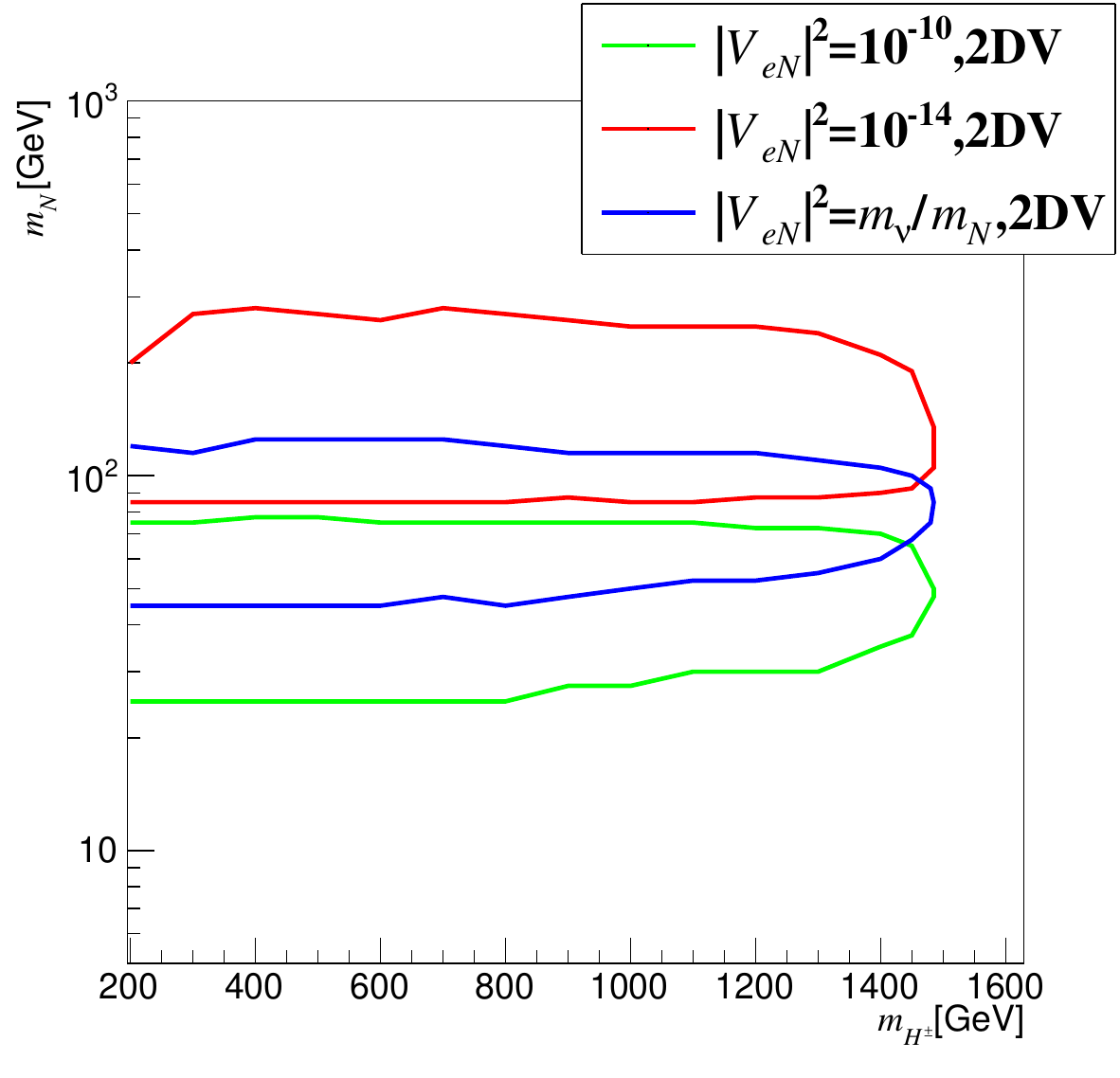}
		\includegraphics[width=0.45\linewidth]{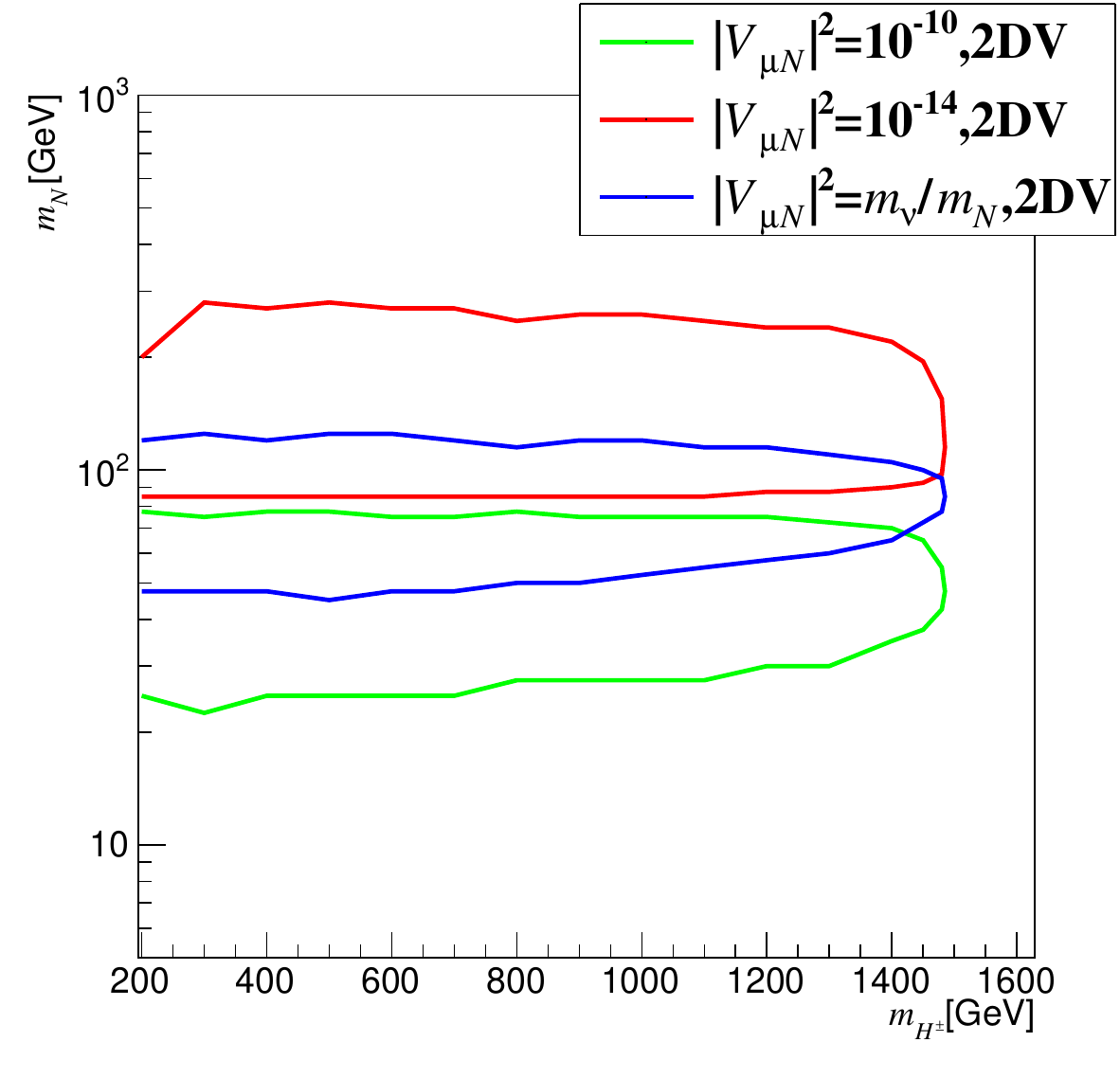}
	\end{center}
	\caption{Same as Fig.~\ref{scan03} but for sensitivity reach of the 3 TeV CLIC with luminosity $\mathcal{L} = 5~{\rm ab}^{-1}$. }
	\label{scan06}
\end{figure}

To obtain the limitation of CLIC on $\nu$2HDM, we then scan the parameter space in the  $m_{H^{\pm}}-m_{N}$ panel with the mixing parameter $|V_{\ell N}|^2=10^{-10}$ and $|V_{\ell N}|^2=10^{-14}$ separately. The results are shown in Figure~\ref{scan06}. Both the one DV and two DVs signature could probe $m_{H^{+}}\lesssim1490$ GeV, which are close to the threshold of $H^+H^-$ production at the 3 TeV CLIC. For $|V_{\ell N}|^2=10^{-10}(10^{-14})$, the promising region of HNL is $m_N\in[15,80]$~GeV ($m_N\in[50,400]$~GeV), with little dependence on the scalar mass $m_{H^+}$. This is because the cross section of $H^+H^-$ at CLIC is nearly a constant for $m_{H^+}$ below the TeV scale. Compared with LHC, the CLIC has a larger sensitive region for $m_{H^+}$ above TeV, mainly due to a larger cross section of $H^+H^-$. For the two DVs signature, the sensitive regions are smaller than the one DV signature. And the sensitive regions for $|V_{\ell N}|^2=10^{-10}$ and $|V_{\ell N}|^2=10^{-14}$ have no overlap. For the seesaw induced mixing $|V_{\ell N}|^2=m_\nu/m_N$, the promising region is about $m_N\in[20,160]$ GeV for one DV signal and $m_N\in[50,110]$ GeV for two DVs signal, with little dependence on the charged scalar mass.

\section{Conclusion}\label{SEC:CL}

The neutrinophilic two Higgs doublet model can naturally explain the tiny neutrino masses with TeV scale heavy neutral leptons. Under a global $U(1)_L$ symmetry, the new Higgs doublet $\Phi_\nu$ carries lepton number $L_{\Phi_\nu}=-1$, while the heavy neutral leptons $N$ have $L_N=0$. Such charge assignment allows the new Yukawa interaction $\bar{L}\tilde{\Phi}_\nu N$, which induces neutrino masses via seesaw mechanism with MeV scale VEV $v_\nu$.

With proper mixing parameter $V_{\ell N}$, the heavy neutral lepton $N$ becomes long-lived, which leads to displaced vertex signature at colliders. In this paper, we consider the displaced vertex signal of the heavy neutral lepton from the neutrinophilic Higgs doublet decay. Compared with the current experimental searches via the $W^\pm\to \ell^\pm N$ channel, the neutrinophilic scalar decay channels as $H^\pm\to \ell^\pm N$ are not suppressed by the small mixing parameter, which makes the new channel more promising at colliders.

In this paper, we perform a detailed simulation of the displaced vertex signature at the 14 TeV HL-LHC in the pure electron or muon mixing pattern. For the 14 TeV HL-LHC with an integrated luminosity of $\mathcal{L} = 3~{\rm ab}^{-1}$, we focus on the current inner tracker DV searches.  According to the simulations, the one DV signature is promising to probe the parameter space with $|V_{\ell N}|^2\gtrsim10^{-19}$ and $m_N<m_{H^+}$, which is about ten orders of magnitudes smaller than the $W^\pm\to \ell^\pm N$ channel. Meanwhile, the promising regions of two DVs signal are slightly smaller than the one DV signal due to more tight selection cuts. We also find that both the one DV and two DVs signature could detect the seesaw predicted mixing $|V_{\ell N}|^2\sim m_\nu/m_N$ for certain $m_N$. The sensitive region of $V_{\ell N}$ heavily depends on the HNL mass $m_N$, i.e., a larger $m_N$ usually needs a smaller $V_{\ell N}$ to satisfy the DV cuts. The DV signatures also depend on the charged scalar mass $m_{H^+}$. At the 14 TeV LHC, we may probe $m_{H^+}\lesssim1200$ GeV via the one DV signal and $m_{H^+}\lesssim1100$ GeV via the two DVs signal. For mixing parameter larger than seesaw value $|V_{\ell N}|^2\sim m_\nu/m_N$, we may probe $m_N<m_W$. While for mixing parameter smaller than the seesaw value, we may detect $m_N$ up to a few hundred GeV. 

A similar analysis is also performed at the 3 TeV CLIC in searching for long-lived HNL. For light charged scalar $m_{H^+}$ around the electroweak scale, the DV signature is less promising at CLIC than at LHC, because the production cross section of $H^+H^-$ is smaller. But it is more promising at CLIC for the TeV scale $m_{H^+}$. We could probe the charged scalar mass up to the threshold $m_{H^+}< \sqrt{s}/2$ in both one DV and two DVs signatures.

Our results can be seen as a supplement to possible future searches of long-lived HNL, which is quite promising via charged scalar decay $H^\pm\to \ell^\pm N$ on LHC and CLIC experiments.
Furthermore, a better understanding of the backgrounds is also highly desirable. For example, if we can reduce the cut on the reconstructed displaced vertex mass and keep the background  clear from the decay of $B$ hadrons, HNL with smaller masses would have a greater chance of being discovered.

\section{Acknowledgments}

This work is supported by the National Natural Science Foundation of China under Grant No. 12375074, 12175115, and 11805081, Natural Science Foundation of Shandong Province under Grant No. ZR2019QA021 and ZR2022MA056, the Open Project of Guangxi Key Laboratory of Nuclear Physics and Nuclear Technology under Grant No. NLK2021-07.


\end{document}